\documentclass{rmf-d}
\usepackage{nopageno,multicol,times,epsf,amsmath,amssymb} 
\usepackage[english]{babel}  
\usepackage{caption2}
\usepackage{graphicx,float,bm}
\usepackage{capt-of}
\usepackage[colorlinks, citecolor=red, linkcolor=blue]{hyperref}
\usepackage{multirow}
\usepackage[numbers]{natbib}
\usepackage{ulem}
\usepackage{tabularx}
\usepackage{adjustbox}
\usepackage{xcolor}
\usepackage{siunitx}
\usepackage[T1]{fontenc}
\def\rmfcornisa{\hfill\rmf\ {\bf }  
\hfill }

\DeclareSIUnit[number-unit-product = {}]{\inchQ}{\textquotedbl}

\DeclareSIUnit[number-unit-product = {\thinspace}]{\inch}{in}

\DeclareMathOperator\erf{erf}
\DeclareMathOperator\erfc{erfc}

\newcommand{\daniel}[1]{{\color{blue} #1}}
\def\rmfcintilla{{\it } {\bf } }

\clearpage \rmfcaptionstyle 
\begin{document}
\title{An experimental setup to generate narrowband bi-photons via four-wave mixing in cold atoms}

\author{N. Arias-Téllez, I. F. Ángeles-Aguillón, D. Martínez-Cara,  A. Martínez-Vallejo, L. Y. Villegas-Aguilar, L. A. Mendoza-López, Y. M. Torres, R. A. Gutiérrez-Arenas, R. Jáuregui  and  D. Sahagún Sánchez}
\address{Instituto de Física, Universidad Nacional Autónoma de México, Circuito de la Investigación Científica s/n, Ciudad Universitaria 04510 Ciudad de México, México}

\author{I. Pérez Castillo}
\address{Departamento de F\'isica, Universidad Autónoma Metropolitana-Iztapalapa, San Rafael Atlixco 186, Ciudad de México 09340, Mexico}

\author{A. Cerè}
\address{Center for Quantum Technologies, 3 Science Drive 2, Singapore 117543}

\maketitle 



\begin{abstract}
We present our recently-built experimental setup designed to generate near-infrared and narrow-band correlated photon pairs by inducing four-wave mixing in a cold gas of $^{87}$Rb atoms confined in a magneto-optical trap. The experimental setup and its automation and control approach are described in detail. A characterization of the optical density of the atomic ensemble as well as the basic statistical measurements of the generated light are reported. The non-classical nature of the photons pairs is confirmed by observing a violation of  Cauchy-Schwarz inequality  by a factor of 5.6 $\times 10^5$ in a Hanbury Brown -- Twiss interferometer. A $1/e$ coherence time for the heralded, idler photons of $4.4 \pm 0.1$\,ns is estimated from our observations. We are able to achieve a value of $10^{4}$\,s$^{-1}$ pair-detection-rate, which results in a spectral brightness of 280\,(MHz s)$^{-1}$.
The combination of high brightness and narrow-band spectrum makes this photon-pair source a viable tool in fundamental studies of quantum states and opens the door to use them in quantum technologies.

\end{abstract}


\begin{multicols}{2}

\section{Introduction}
\label{sec:intro}
Since the 1970s sources of entangled photon pairs are an important pillar in quantum optics. On the one hand they enable to probe the foundations of quantum mechanics and, on the other, they are a cornerstone in the development of novel quantum technologies. Time-correlated photon pairs were firstly generated by inducing spontaneous parametric down conversion (SPDC) in non-linear crystals~\cite{burnham1970observation}. This was subsequently followed by observing quantum entanglement for the first time in photons emitted by atomic gases subjected to a four-wave mixing (FWM) process ~\cite{freedman1972experimental}. In addition to serving a broad span of basic research during several decades, experimental methods based on these two non-linear processes play an important role in the \textit{second quantum revolution}, the on-going development of technologies that is set to change -- even -- daily life paradigms~\cite{Dowling:2003eia}.

With SPDC, researchers have invented techniques yielding an outstanding feedback between science and technology. Several of those methods are now the platform for both exploring the laws of quantum physics and for developing applications of quantum optics such as cryptography and metrology~\cite{Couteau:2018gv}. The role of FWM as a main character in quantum optics and its applications is spread in history. After subsequent experiments establishing methods for testing the Bell inequalities~\cite{aspect1982experimental} this non-linear process attracted back major attention after the proposal by Duan, Lukin, Cirac and Zoller for building a quantum repeater based on atomic gases \cite{Duan:2001dz}. Such repeater has not yet been achieved, however, the challenging quest for it has engaged researchers in the development of long-lived and efficient quantum memories~\cite{cho2016highly, guo2019high}, and the generation of photon pairs that can -- in principle -- be tailored by means of the exquisite control over the atomic states provided by precision spectroscopy. 

\begingroup
    \centering
    \includegraphics[width=0.6\columnwidth]{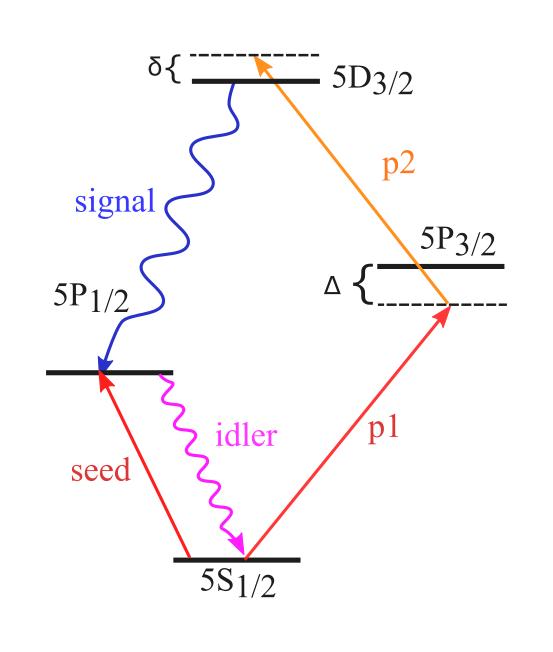}
    \captionof{figure}{Simplified schematics of the FWM diamond configuration implemented in our experiments. It involves a ladder transition followed by a cascade decay in the Rb$^87$ electronic structure: $p1$ represents a laser beam stabilized to its 5S$_{1/2} \rightarrow$ 5P$_{3/2}$ and $p2$ depicts resonant to the 5P$_{3/2} \rightarrow$ 5D$_{3/2}$  transition. Both detunings $\Delta$ and $\delta$ are important parameters for the non-linear process.  The signal photons are emitted when atoms decay from the 5D$_{3/2}$ to the 5P$_{1/2}$ state; idler photons are parametric decay happening generated by the 5P$_{1/2} \rightarrow$ 5S$_{1/2}$ decay. A seed laser,  tuned to the  5S$_{1/2}$ $\rightarrow$ 5P$_{1/2}$ transition, is used for amplifying the FWM signal in the pre-alignment procedure. However, it is switched off during experiments as explained in Section \ref{sec:setup}.      }\label{fig:diamante}
\endgroup

Motivated by the creation of hybrid quantum systems~\cite{Wallquist:2009efa,Kurizki:2015ew} time-correlated photon pairs were generated from cold atoms in a seminal work by Kuzmich {\it et al.}~\cite{Kuzmich:2003fe}. Their potential as a narrowband source of photon pairs suitable for interacting with atoms and thus, for creating quantum systems, was thereafter demonstrated~\cite{Thompson:2006gf}. In this context it is desirable to generate photons that can excite  specific dipole lines connecting predetermined atomic states, and to generate photons in the telecommunications regime. In this way, one would be able to imprint information into a quantum memory and simultaneously send it far away through optical fibers. In principle, this can be achieved with several \textit{diamond} schemes of transitions within the rich electronic structure of the alkali elements~\cite{Chaneliere:2006gx} (see Figure~\ref{fig:diamante}). 

Photons generated by FWM in laser-cooled atoms have naturally a bandwidth comparable to the decay rate of their states (in the order of a few MHz). Using electromagnetically induced transparency it is possible to achieve bandwidths below one MHz~\cite{Du:2008do}.  With the diamond scheme, a bandwidth of 20 MHz has been reached without external intervention~\cite{srivathsan2013narrow}. 
It can be further narrowed below 8 MHz by introducing a resonant cavity~\cite{Seidler:2020jm}. In setups of this kind, it is possible to generate photonic Bell states; entanglement of linear momentum~\cite{Willis:2011ta,Gulati:2015ee} and orbital angular momentum~\cite{Zhao:2019ix} has also been observed between the signal and idler counterparts. These states are directly related to the atomic degrees of freedom and the dynamics induced by the FWM process. Therefore, they are  controllable with the characteristics of the laser beams pumping the non-linear process. 

Within this context we present our experimental setup for generating photon pairs from cold atoms as well as measurements of the  coherence of the photons to show that it is, indeed, a narrowband and bright source of non-classical light. The remainder of this article is organized as follows. In Section \ref{sec:FWM_diamond} we describe a theoretical framework of FWM in atoms, from which the phase matching conditions are derived. This is followed by a model of the cascade decay  useful in the description of the generated bi-photon state (Section \ref{sec:theory_coherence}). Next, in  Section \ref{sec:setup}, we describe in detail the experimental setup. There, we emphasize basic concepts of the applied methods, and describe technical aspects particular to our machine and its essential experimental protocols. In Section \ref{sec:Statistical_measurements} we show the experimental results and their analysis. From their second-order crossed-correlation function we estimate the coherence time of the idler as heralded by the signal photons. With this time we compute the spectral brightness of our source for typical experimental conditions. Following that, we verify the chaotic time statistics of the signal and idler photons in terms of the corresponding auto-correlation measurements. We then use both results to evaluate the Cauchy-Schwartz inequality, showing a strong violation. Finally, in Section \ref{sec:conclusion_outlook}, we conclude the reported work and give an outlook toward our research interests.

\section{Four-wave mixing in atoms}
\label{sec:FWM_diamond}

Four-wave mixing is possible in center-symmetric media, such as an atomic gas. Figure \ref{fig:fwm_mot} illustrates its implementation in a cold atomic gas within a magneto-optical trap (MOT). The pump beams $p1$ and $p2$ target the cloud of cold atoms. Their field overlap in the same volume where the signal (\textit{s}) and idler (\textit{i}) photons are generated and eventually detected with avalanche photodiodes.   

\begingroup
    \centering
    \includegraphics[width=0.9\columnwidth]{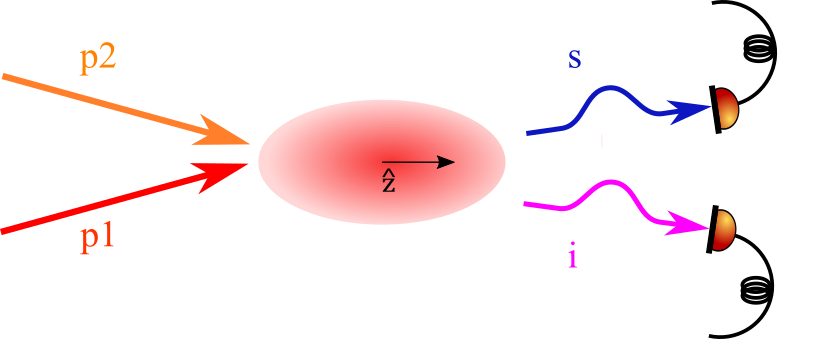}
    \captionof{figure}{Generic schematics of FWM in cold atomic gases. Two pump, $p1$ and $p2$, are overlapped in the center of a MOT. A signal (\textit{s}) and idler (\textit{i}) photon pair is generated inside the interaction volume. }\label{fig:fwm_mot}
\endgroup

The stimulated atomic transitions come along with isotropic spontaneous parametric emission of photons. However, a proper phase relationship between the interacting waves guarantees constructive addition of the amplitude contributions from different locations and emission times inside the atomic gas. Both processes can be described in terms of the
electric susceptibilities: isotropic spontaneous emission in terms of first order susceptibility $\chi^{(1)}$ and FWM in terms of the third order susceptibility  $\chi^{(3)}$.
 The later relates the polarization of atoms with  the involved electric fields as
$$\mathbf{P}_j^{(+)}(\omega_{s,i})=$$
\begin{align}\label{eq:x3_fwm}
\epsilon_0\sum_{klm}\chi_{jklm}^{(3)}(\mathbf{E}^{(+)}_{p1})_k(\mathbf{E}^{(+)}_{p2})_l(\mathbf{\hat{E}}^{(-)}_{i,s})_m
\end{align}
inside the material. In this case the polarization acts as a source of quantum light at the signal or idler angular frequencies ($\omega_{s}$ or $\omega_{i}$).  In Equation~ (\ref{eq:x3_fwm}), $\epsilon_0$ is the electric permittivity of vacuum. The two pump electric fields $\mathbf{E}_{p1}$ and $\mathbf{E}_{p2}$ are  written in terms of  plane wave modes
\begin{eqnarray}
\mathbf{E}_{p}&=&\frac{1}{2}\left(\mathbf{E}^{(+)}_{p}+\mathbf{E}^{(-)}_{p}\right) \nonumber\\
    \mathbf{E}^{(+)}_{p}(\mathbf{r},t)&=& \mathbf{E}_{p}e^{i(\mathbf{k}_{p}\cdot \mathbf{r}-\omega_{p}t)} \\
\mathbf{E}^{(-)}_{p}(\mathbf{r},t)&=& \mathbf{E}_{p}^*e^{-i(\mathbf{k}_{p}\cdot \mathbf{r}-\omega_{p}t)}       \nonumber
\end{eqnarray}
with $p=p1$, $p2$; the angular frequencies $\omega_{p1}$ and $\omega_{p2}$, and the wave vectors $\mathbf{k}_{p1}$ and $\mathbf{k}_{p2}$. The pump electric fields are well represented by classical fields because they are laser beams. However, the signal an idler fields must  be described using the quantum theory,
$$ 
\mathbf{\hat{E}}^{(+)}_{s,i}(\mathbf{r},t)=
\frac{1}{2\pi}\int d\omega_{s,i} d\mathbf{k}\sqrt{\frac{2\hbar\omega_{s,i}}{c\epsilon_0A}}   \hat{a}_{s,i}(\omega_{s,i},\mathbf{k}_{s,i},\boldsymbol{\epsilon}_{s,i})$$
\begin{equation}\label{eq:qfields}
\boldsymbol{\epsilon}_{s,i}(\mathbf{k}_{s,i}) e^{i\mathbf{k}_{s,i}\cdot\mathbf{r}-\omega_{s,i} t)}
\delta(\omega_{s,i}-\vert{\mathbf{k}}_{s,i}\vert),
\end{equation}
where $\boldsymbol{\epsilon}$ denotes the polarization vector and  $A$  an  effective cross-section area introduced for quantization purposes. The creation  and
annihilation operators,  $\hat{a}^\dagger_{s,i}$  and $\hat{a}_{s,i}$, satisfy the standard commutation relations for all the variables determining the mode,  $$\left[\hat{a}_{s,i}(\omega,\mathbf{k},\boldsymbol{\epsilon}),\hat{a}_{s,i}^\dagger(\omega^\prime, \mathbf{k}^\prime,\boldsymbol{\epsilon}^\prime)\right]=\delta_{\epsilon,\epsilon^\prime}\delta(\omega - \omega ^\prime)\delta\left(\mathbf{k} - \mathbf{k}^\prime\right);$$
$\mathbf{\hat{E}}^{(-)}_{s,i}$ is the Hermitian conjugate of the operator $\mathbf{\hat{E}}^{(+)}_{s,i}$.

In the interaction picture, the  relevant effective Hamiltonian of the atomic system, whose properties are encoded in the susceptibility $\chi^{(3)}$, and the four electric fields  is \cite{wen2006transverse}
$$    \hat{\mathbf{H}}_I=$$
\begin{equation} \label{eq:Hamiltonian_i}
\frac{\varepsilon_0}{4}\sum_{jklm}\int_Vd\mathbf{r}\chi_{jklm}^{(3)}(\mathbf{E}^{(+)}_{p1})_k(\mathbf{E}^{(+)}_{p2})_l(\hat{\mathbf{E}}^{(-)}_s)_m(\hat{\mathbf{E}}^{(-)}_i)_j + \textrm{H. c.},
\end{equation}
here $d\mathbf{r}$ denotes the differential of the volume $V$ where the four fields overlap. Considering $p1$ and $p2$ as co-propagating plane waves simplifies the mathematical visualization of the physical process and,  even more important, resembles the system created in our setup. By integrating the Hamiltonian (\ref{eq:Hamiltonian_i}) over the complete space,  one finds two  phase-matching conditions, that can be interpreted as conservation rules,
\begin{equation}\label{eq:k_conservation}
    \mathbf{k}_{p1} +  \mathbf{k}_{p2} =  \mathbf{k}_{s}+ \mathbf{k}_{i},
\end{equation}
for linear momentum and 
\begin{equation}\label{eq:omega_conservation}
    \omega_{p1}+\omega_{p2} = \omega_s + \omega_i,
\end{equation}
for energy. Equation (\ref{eq:k_conservation}) relates the direction of the generated photons with that of the pump beams. In this case all are co-linear.

\section{Coherence of the generated quantum states}
\label{sec:theory_coherence}

The time statistics of the photon pairs can be evaluated with the second-order correlation function 
\begin{align}\label{eq:glauber}
    \mathcal{G}^{(2)}(t, t+\Delta t) &= 
    || \mathbf{\hat{E}}^{(+)}_s(\textbf{r}_1,t)\mathbf{\hat{E}}^{(+)}_i(\textbf{r}_2,t+\Delta t)|\psi\rangle ||^2
\end{align}
with  $\textbf{r}_{1,2}$ the positions where they are detected, and $t$ and $t+\Delta t$ the detection times. If a steady state can be reached, 
\begin{equation}
{\lim_{t \rightarrow \infty}\mathcal{G}^{(2)}(t, t+\Delta t) = G^{(2)}(\Delta t)}.
\end{equation}

\subsection{Heralding effect}
\label{sub:heralding_effect}
For quantum information applications, it is important to know when a single photon is likely to be detected. Photons emitted in cascade facilitate this due to the intrinsic delay between photons of each pair. This can be evaluated from their second order correlation function which can be calculated by solving the  evolution equation of the atom, idler and signal electromagnetic field state,
\begin{equation}\label{eq:shr}
\frac{d}{dt}|\psi(t)\rangle=-\frac{i}{\hbar}\tilde{H}_I|\psi(t)\rangle.
\end{equation}
 Instead of solving the Maxwell-Bloch equations, we assume a simplified scheme where just the atomic states participating in the cascade transitions are involved, so that the relevant atom-electromagnetic field state can be expressed as 
\begin{align}\label{eq:psi_state}
    |\psi(t)\rangle =& c_\alpha(t)|\alpha,0_k,0_q\rangle + \sum_kc_{\beta k}(t)|\beta,1_k,0_q\rangle\\ &+ \sum_{k,q}c_{\gamma kq}(t)|\gamma,1_k,1_q\rangle.\nonumber
    \end{align}
For the FWM configuration in Figure~\ref{fig:diamante},  the first term of this equation corresponds to the  atomic second-excited state  $|\alpha\rangle = 5D_{3/2}$ with zero signal or idler photons, the second term is  linked to the atomic first excited state  $|\beta\rangle = 5P_{1/2}$ with one signal photon occupying one vacuum mode $k$ and zero idler photons, and the last term stands for the ground state  $|\gamma\rangle = 5S_{1/2}$ with the presence  of one signal and one idler photon in the mode $q$. The effective Hamiltonian in the interaction picture under the rotating-wave approximation is \cite{scully1999quantum}
\begin{align}\label{eq:Hi}
\tilde{H}_I=\hbar\sum_k\left(g_{\alpha \beta}^{sk}\sigma^{(1)}_+\hat{a}_{ik} e^{i(\omega_{\alpha\beta}-\omega_k)t}+H. c\right)+ \\ \hbar\sum_q\left(g_{\beta \gamma}^{iq}\sigma^{(2)}_+\hat{a}_{sq} e^{i(\omega_{\beta\gamma}-\omega_q)t}+H. c\right), \nonumber
\end{align}
where $\omega_{if}$ is the frequency corresponding to the energy separation between states for each  $|i\rangle\rightarrow|j\rangle$ decay and, the raising operators  are $\sigma_+^{(1)}=|\alpha\rangle\langle \beta|$ and $\sigma_+^{(2)}=|\beta\rangle\langle \gamma|$. In Equation~(\ref{eq:Hi}) the coupling constants,
\begin{equation}\label{eq:gij}
g_{ij}^{lk}=\sqrt{\frac{\hbar\omega_k}{2\epsilon_0V}}\textbf{d}_{ij}\cdot\boldsymbol{\epsilon}_{k}(\textbf{k}) e^{i\textbf{k}\cdot\textbf{r}},
\end{equation}
are the analogous to  Rabi frequencies for single photons; they are defined in terms of the dipole transition matrix element $\textbf{d}_{ij}$.

One can solve for the coefficients $c(t)$ of the state $|\psi(t)\rangle$ by inserting Eqs.~ (\ref{eq:psi_state}-\ref{eq:Hi}) into Equation~(\ref{eq:shr}). Introducing the $\Gamma_\alpha$ and $\Gamma_\beta$ the decay rates of the states $|\alpha\rangle$ and $|\beta\rangle$, the system of differential equations 
\begin{align}\label{eq:c_eqtns}
\Dot{c}_\alpha(t) & =-\frac{\Gamma_\alpha}{2}c_\alpha(t), \nonumber\\ 
\Dot{c}_{\beta k}(t) & = - ig_{\alpha\beta}^{sk}e^{i(\omega_{\alpha\beta}-\omega_k)t}c_a(t)-\frac{\Gamma_\beta}{2}c_{\beta k}(t),\\
\Dot{c}_{\gamma k q}(t) & = -i g_{\beta\gamma}^{iq} e^{i(\omega_{\beta\gamma}-\omega_q)t}c_{\beta k}(t),\nonumber
\end{align}
is obtained.
The decay rate $\Gamma_\beta=1/(2\pi\tau_c)$ determines a coherence time for the heralded idler photons. The coefficients $c(t)$  are constrained to the initial conditions $c_\alpha(0)=1$ and  $c_{\beta k}(0)=c_{\gamma kq}(0)=0$ in order to resemble a cascade decay.



From the first equation $$c_\alpha (t)= \exp{(-\Gamma_\alpha t/2)}$$ is obtained trivially. The second coefficient
\begin{equation}
    c_{\beta k}(t) = ig_{\alpha \beta}^{sk}\frac{e^{[i(\omega_k-\omega_{\alpha\beta})-\Gamma_\alpha/2]t}-e{^{-\Gamma_\beta t/2}}}{i(\omega_k-\omega_{\alpha\beta})-(\Gamma_\alpha-\Gamma_\beta)}.
\end{equation}
is readily derived by inserting this solution into the corresponding equation and integrating; plugging this solution of $ c_{\beta k}(t)$ into the third differential equation yields the long time limit 
\begin{align}\label{eq:cgamma_kq}
    c_{\gamma kq}(\infty) &= \frac{g_{\alpha\beta}^{sk}}{i(\omega_k+\omega_q-\omega_{\alpha\beta}-\omega_{\beta\gamma})-\Gamma_\alpha/2)} \\ &\times  \frac{g_{\beta\gamma}^{iq}}{i(\omega_q-\omega_{\beta\gamma})-\Gamma_\beta/2}.\nonumber
\end{align}
The long term limit for the other to coefficients is $c_\alpha(\infty)=c_{\beta k}(\infty)=0$. 
Thus, these equations yield Lorentzian profiles for the two decays.

The cross-correlation function for the cascade decay can now be evaluated for the asymptotic steady state
\begin{equation}
    |\psi(\infty)\rangle=\sum_{kq}c_{\gamma kq}(\infty)|\gamma,1_k,1_q\rangle,
\end{equation}
and the signal and idler fields shown in Equation~(\ref{eq:qfields}). It can be calculated from  
\begin{equation}\label{eq:G2_si}
    G_{si}^{(2)}(\Delta t)=\langle \psi(0)|\mathbf{\hat{E}}^{(+)}_s\mathbf{\hat{E}}^{(+)}_i|\psi(\infty)\rangle,
\end{equation}
which is equivalent to Equation~(\ref{eq:glauber}). In the limit for a continuum vacuum spectrum it yields a double integral that can be written in terms of wavevectors $\mathbf{k}$ and $\mathbf{q}$. After solving its angular part in spherical coordinates 
\begin{align}\label{eq:G_SI}
    G_{si}^{(2)}(\Delta t) & = G_0
 \int_0^{\infty}\int_0^{\infty} dk dq \frac{e^{-ckt_1}(e^{-ik\Delta r_1}-e^{ik\Delta r_1})}{[i(ck+cq-\omega_{\alpha\gamma})-\Gamma_\alpha/2)]} \\ 
 &\times \frac{e^{-ckt_2}(e^{-ik\Delta r_2}-e^{ik\Delta r_2})}{[i(cq-\omega_{\beta\gamma})-\Gamma_b/2)]}. \nonumber
\end{align}
with the constant $G_{0}$ defined as
\begin{equation}
    G_0= \frac{(2\pi)^2k_0q_0g_{\alpha k}g_{\beta q}}{\Delta r_1\Delta r_2},
\end{equation}
where $k_0$ and $q_0$ are constants from the angular integration. Since its integrand is a product of two Lorentzians the lower limit of the integral can be  extended to $-\infty$ for both variables. Thus the residue theorem of Cauchy can be applied, leading to
\begin{align}
   & G_{SI}^{(2)}(\Delta t)= 2\pi^2G_0 e^{-(i\omega_{\alpha\beta}+\Gamma_\alpha/2)t_s} \\
   &  \times \Theta[t_s] e^{-(i\omega_{\beta\gamma}+\Gamma_\beta/2)(t_i-t_s)}  \Theta[t_i-t_s]. \nonumber
\end{align}
where we define $t_s=t_1-\Delta r_1/c$ and $t_i=t_2-\Delta r_2/c$. In our case $\Gamma_{\beta}=36$ MHz and $\Gamma_{\alpha}=0.6$ MHz; $G_{SI}^{(2)}(\tau)$ is dominated by the decay of state $|\beta\rangle$. The detection time of the signal photon $t_s$ is always shorter than the difference in detection time between the idler and signal $t_i-t_s$, so the product of the two Heaviside step $\Theta$ functions simplifies to
\begin{equation}\label{eq:G_2_double_decay}
    G_{SI}^{(2)}(\Delta t)=2\pi^2G_{0}e^{-\Gamma_{\beta}\Delta t/2} \Theta(\Delta t);
\end{equation}
defining $\Delta t=t_i-t_s$. Equation (\ref{eq:G_2_double_decay}) states that $ G_{SI}^{(2)}(\Delta t)$ has a sudden rise to its maximum value followed by an exponential decay with a rate  $\Gamma_{\beta}$, equal to {\color{red} $2\pi/\tau_c$}. In this case $\tau_c$ is the coherence time of the idler photons.  The same result can be found with a treatment based on Green functions \cite{kocher1971time}.  

Note that this treatment involves a single atom, it does not account for collective effects. However,  corrections to this scheme can be estimated from effective decay rates \cite{stroud1972superradiant} as described in Section \ref{sub:Optical-density}.


\section{Experimental setup}
\label{sec:setup}

Two laser frequencies are required for cooling atoms and another three different frequencies are employed to induce FWM in the atoms. All of them are stabilized by common saturation absorption spectroscopy techniques. However, the optical reference for $p2$ is given by a two-photon spectroscopy arrangement that is relatively simple, \daniel{yet} it has not been reported to the best of our knowledge. The core of this setup is a standard MOT of $^{87}$Rb atoms. It is created inside an almost portable vacuum system\footnote{It has an approximate weight of 5 kg and a  volume of about $50\times 30\times 20$\,cm$^{-3}$; the vacuum quality has shown to be resilient to power cuts of the pump for longer than a few days, which makes the whole system suitable for transportation.  }. This MOT confines typically over 10$^7$ atoms with an optical density ($OD$) that can be chosen to take any value between 5 and 22. The control and data acquisition systems are a combination of commercial equipment with home built electronics, controlled by our purpose-developed software. Four-wave mixing is implemented in a pulsed mode that optimizes the coincidence rate of the photon pairs. The photon collection optics has been arranged with two Hanbury-Brown-Twiss interferometers and with two setups for measuring polarization correlations between the signal and idler photons.

\subsection{Laser system}
\label{sub:laser_system}

As shown in Figure~\ref{fig:lasers} the laser system is built in a modular fashion. Each gray box represents an optical arrangement that has been setup on a breadboard made of granite with the optical posts attached to it with instant glue. As reported in~\cite{sahagun2013simple}, we found this to be a convenient solution because it allows building compact optical setups that remain stable for years\footnote{It is desirable to form an evenly thin adhesive layer between post and granite. For this reason the Cretan Matterwaves Group chose Loctite~401 -- the less viscous of these cyanoacrylate-based glues~\cite{sahagun2013simple}. We opted for Loctite~495 instead since the number 401  was not  domestically available.}. The boxes with solid borders represent setups of spectroscopy; boxes with dashed borders are optical arrangements that distribute light. The five frequencies required for inducing FWM in the MOT are generated by three commercial lasers (Moglabs: LDL and CEL002) and one home-built extended cavity diode laser (ECDL).

The master laser is located at the left, top corner of Figure~\ref{fig:lasers}. It emits light locked to the crossover between the 5S$_{1/2}$, F = 2 $\rightarrow$ 5P$_{3/2}$, F' = 3 and the 5S$_{1/2}$, F = 2 $\rightarrow$ 5P$_{3/2}$, F' = 2 transitions to three different points around the setup. About one third of it (20 mW)  seeds a tapered amplifier (TA) that generates up to 2 Watts of light that are further tuned with AOMs for laser cooling and for measuring the optical density $OD$. On the AOM board, the amplified light is offset by + 60 MHz with a single-pass OAM. This leaves it -151.6 MHz from the cooling transition. Thus we can optimize the laser cooling through its detuning $\varepsilon$, and implement the $OD$ measurements described in Section \ref{sub:Optical-density} with an appropriate control of the detuning $\Delta$ of its probe beam with an AOM double passes of + 76 MHz in each case. To close the two-level system required for laser cooling the re-pump beam is combined with the cooling light at the polarizing beam splitter right before its output port as depicted in Figure~\ref{fig:lasers}.  The re-pump laser is home-made,  its design was inspired on the Littrow version reported on \cite{sahagun2013simple}\footnote{This laser was developed in collaboration with the Rydberg Quantum Optics Laboratory of Institute of Physics, UNAM. We made a few adaptations to replace some of the custom-machined parts with readily available off-the-bench components. It has been working almost non-stop for about three years with an excellent stability performance. Moreover we are building a low noise current supply with components that are readily acquirable in Mexico. We will make both the ECDL and the current supply designs available in a future publication.}.  
Clockwise, around the master laser in Figure~\ref{fig:lasers}, its second output  supplies $p1$ light to the AOM board. In order to avoid incoherent scattering from the first step of the FWM it is convenient to set a large detuning $\Delta$ (Figure~\ref{fig:diamante}). Its optimal value was found around - 60\,MHz in our experiments. This is achieved by bypassing the offset AOM with the $p1$ beam, which is delivered to the FWM board after double passing its AOM. To complete the FWM frequencies we have the seed and the $p2$ lasers. The former is  stabilized to the 5S$_{1/2}$, F = 2 $\rightarrow$ 5P$_{1/2}$, F' = 2 transition with standard saturated absorption spectroscopy. As described in Section \ref{sub:pumping}, it is used for alignment purposes without further frequency tuning.

\end{multicols}
\begingroup
    \centering
    \includegraphics[width=0.8\textwidth]{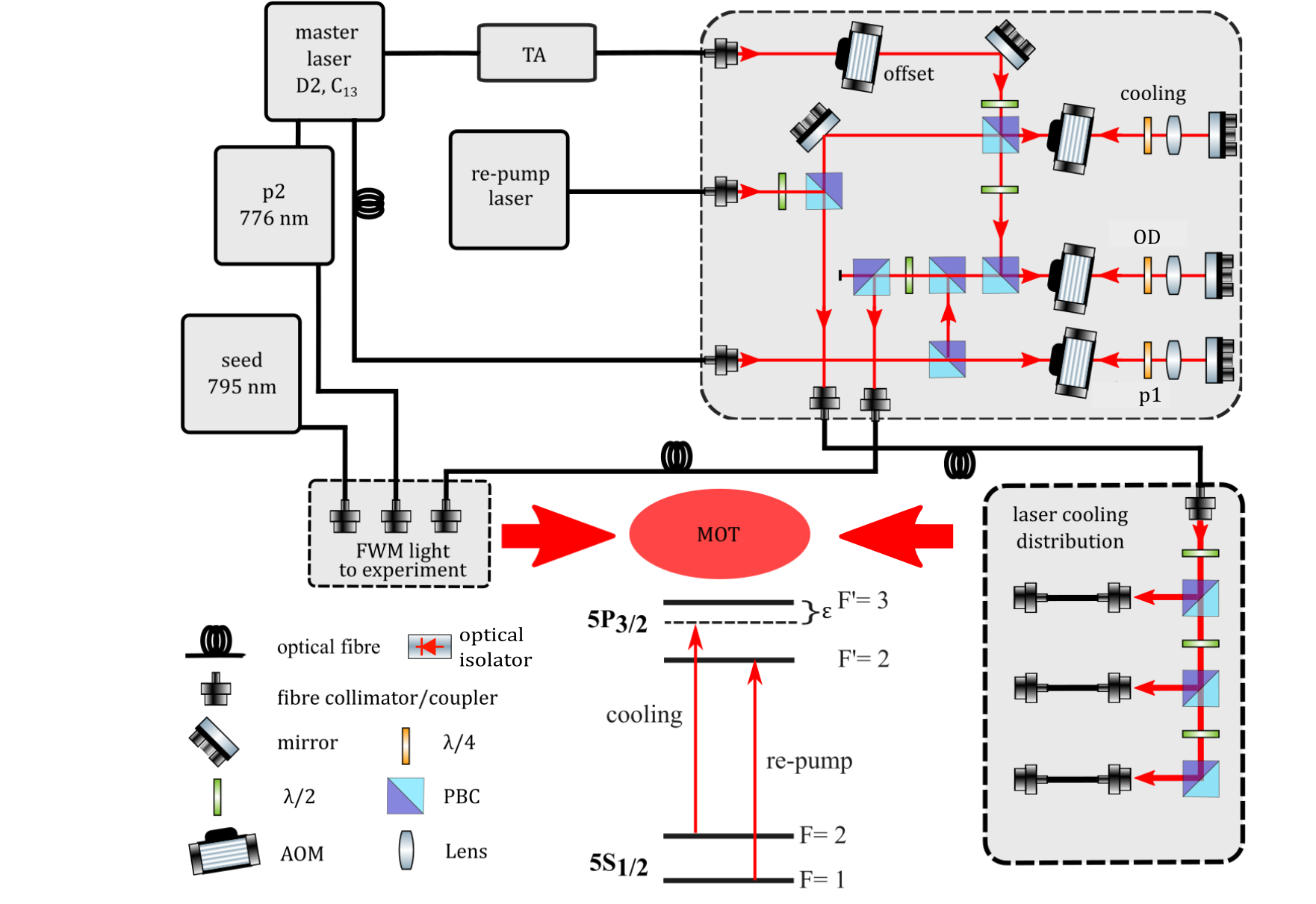}
    \captionof{figure}{Block diagram of our experiment. The master laser is located at the left, top corner. It delivers frequency-stabilized light to seed our tapered amplifier (TA) for laser cooling, a small portion of its light is further tuned to prepare the $p1$ beam and the third output provides the first photon in the two step spectroscopy setup for the $p2$ laser that is described in Section \ref{sub:two_photon}. In addition to the spectroscopy arrangements (gray boxes with solid borders), the light is fine tuned and prepared in polarization on the  distribution boards (gray boxes with dashed borders).    }
    \label{fig:lasers}
\endgroup
\begin{multicols}{2}

\subsubsection{Two-photon spectroscopy}
\label{sub:two_photon} 
 The $p2$ laser, depicted in Figure~\ref{fig:lasers}, is a source of light at 776 nm intended to stimulate the 5P$_{3/2} \rightarrow$ 5D$_{3/2}$ transition in the MOT. This is the second step of a double atomic transition thus, it has a relatively low probability to happen.  A laser stabilization scheme, based on fluorescence lock-in detection, is suggested in Ref.~\cite{mandal2018blue}. Among the standard techniques we found that the method proposed in~\cite{grove1995two} to probe the  5D$_{5/2}$ state in room-temperature atoms by increasing the Rubidium pressure inside the spectroscopy cell, can be adapted to our scheme. Implementing polarization spectroscopy \cite{Harris:2006} yields similar results. However, temperature drifts due to the cell heating makes stabilization hard to establish in the long term. 

Figure~\ref{fig:two_photon} (a) shows a schematics of our two-photon spectroscopy setup. About 3\,mW of the master laser light is delivered to it through the optical fiber IF. The frequency of this beam is shifted by 2 $\times$ 70\,MHz. This compensates the - 60\,MHz detuning on the $p1$ beam from resonance and enables a convenient scanning of $\delta$ with the $p2$ laser. The same AOM modulates the frequency of the master laser by 250\,kHz as required for saturated absorption spectroscopy. This beam is then sent through the cell as the pump with horizontal polarization.

\begingroup
    \centering
    \includegraphics[width=\columnwidth]{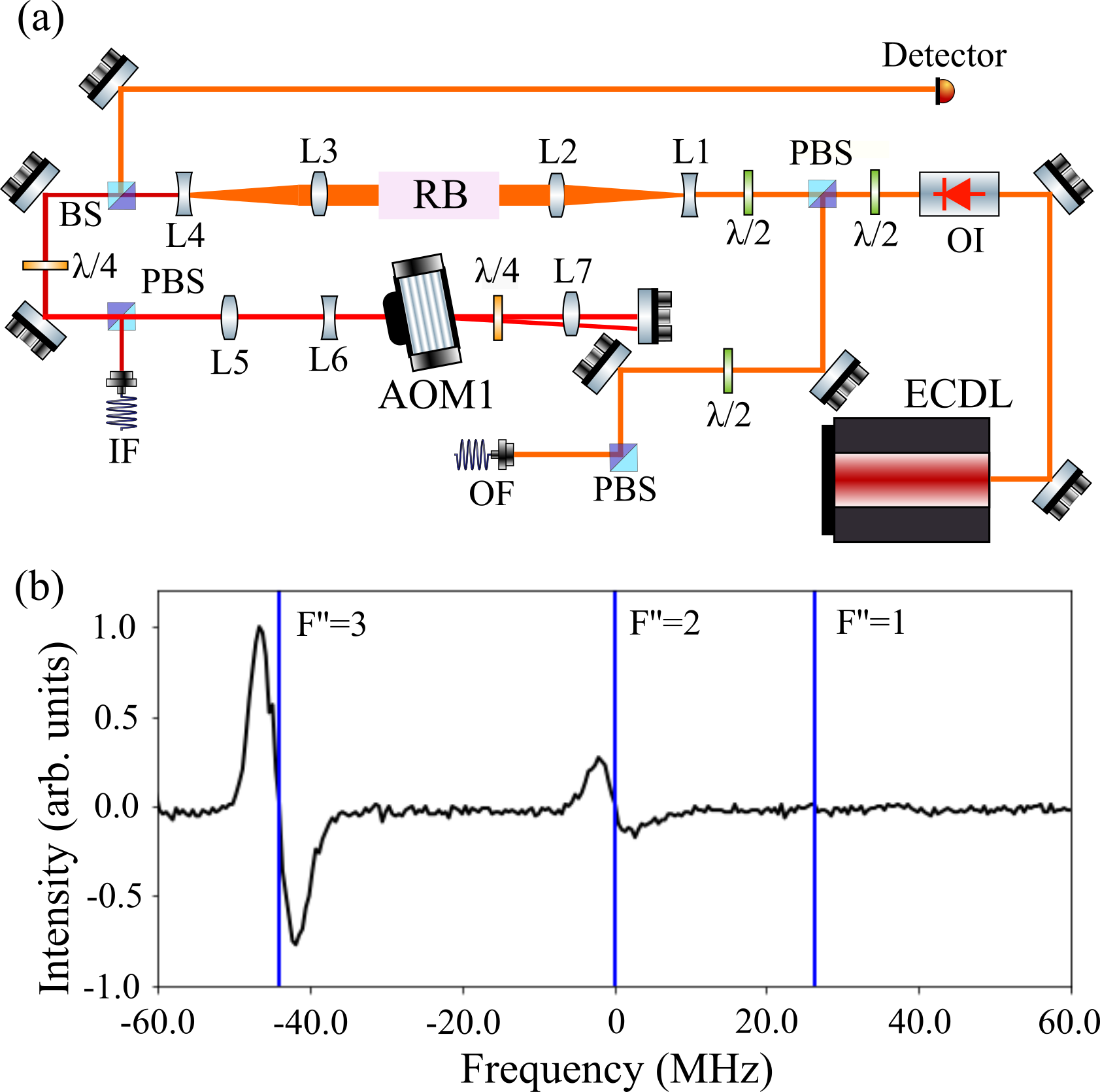}
    \captionof{figure}{Schematics of the two-photon spectroscopy setup (a) and error signal obtained with it (b). The red beam depicts light at 780\,nm for exciting the first step of the ladder; the orange beam represents the second photon, at 776\,nm. IF indicates the input fiber of the 780\,nm beam and OF indicates the output fiber of $p2$ going to the experiment. Optical elements are labeled following the same code as in Figure~\ref{fig:lasers}.}\label{fig:two_photon}
\endgroup

The p2 light is generated by a cateye ECDL \cite{thompson2012narrow} (MOGLABS, CEL002). Its frequency is scanned across all the hyperfine levels of the 5D$_{3/2}$ state. Approximately 3\,mW from this laser enters the spectroscopy cell as the probe beam with linear polarization. Both beams counter-propagate and fully overlap to cancel Doppler broadening and to increase the interaction volume. The later is further incremented by expanding  their $1/e^2$ width from 1 mm to 4 mm, see Figure~\ref{fig:two_photon} (a).

Modulating the frequency of the pump beam also modulates its interaction with atoms inside the spectroscopy cell. This facilitates to derive the saturation signal assisted with a servo-loop circuit and a low-pass filter~\cite{barger1969pressure}, yielding 
a 20\,mV peak-to-peak error signal in this case. This is a factor of 15 smaller than typical traces probing D lines with saturation absorption spectroscopy. Heating the spectroscopy cell enhances the obtained signal, making it suitable for locking a laser with standard electronics. Two power resistors of 10 $\Omega$ carrying 2\,A are attached to the base of the cell. Thus, a stable  temperature close to 70$^{\circ}$C is reached after one hour of heating. Figure~\ref{fig:two_photon} (b) shows a typical error trace that serves for locking the $p2$ laser. The blue lines cross the curve at the three points in which we can lock it. For the experiments reported here we used the $F'=3$ to $F''=3$ of the 5P$_{3/2}, F'  \rightarrow$ 5D$_{3/2}, F''$ transition. If one desires locking the laser to $F'' = 1$ or $2$,  it would be necessary to further increase the temperature of the cell. This laser produces 25\,mW of stabilized light. Its detuning  $\delta$ (Figure~\ref{fig:diamante}) is controlled with the AOM in the range  - 12 to + 12\,MHz with respect to the 5P$_{3/2}, F'=3  \rightarrow$ 5D$_{3/2}, F''=3$ transition.

\subsection{Vacuum system and MOT}
\label{sub:MOT}

As depicted in Figure~\ref{fig:vacuum}, the vacuum system has six elements only. Its simplicity grants  full optical access and enables to reach ultra-high vacuum (UHV). The science cell (1) is an octagon made of quartz with windows anti-reflection coated for 780 nm light. The top and bottom windows have a 1.375\,in diameter whereas the diameter of  the lateral windows is 1\,in. This cell is connected to a CF30 cube that holds all the vacuum system (2). Its vertical branch runs upwards towards a hybrid ion-sublimation pump [NEXTorr D 100-5, (3)]. Lastly (4) is a tee connecting a four-conductor feedthrough (5) and a UHV valve (6). The feedthrough connects independently two Rubidium dispensers and the valve is used for plugging the system to a pumping station for building vacuum from atmospheric pressure. With this system  it is possible to reach pressures below 10$^{-10}$\,Torr without baking. Ultra-high vacuum is not a necessary requirement for creating a MOT. However, in it one can transfer atoms to a dipole or a magnetic trap. With those traps it is possible to change the shape of  the atom cloud. This would open the possibility of modulating the superradiance generated by collective effects throughout the FWM process \cite{jen2012spectral}.

\begingroup
    \centering
    \includegraphics[width=01\columnwidth]{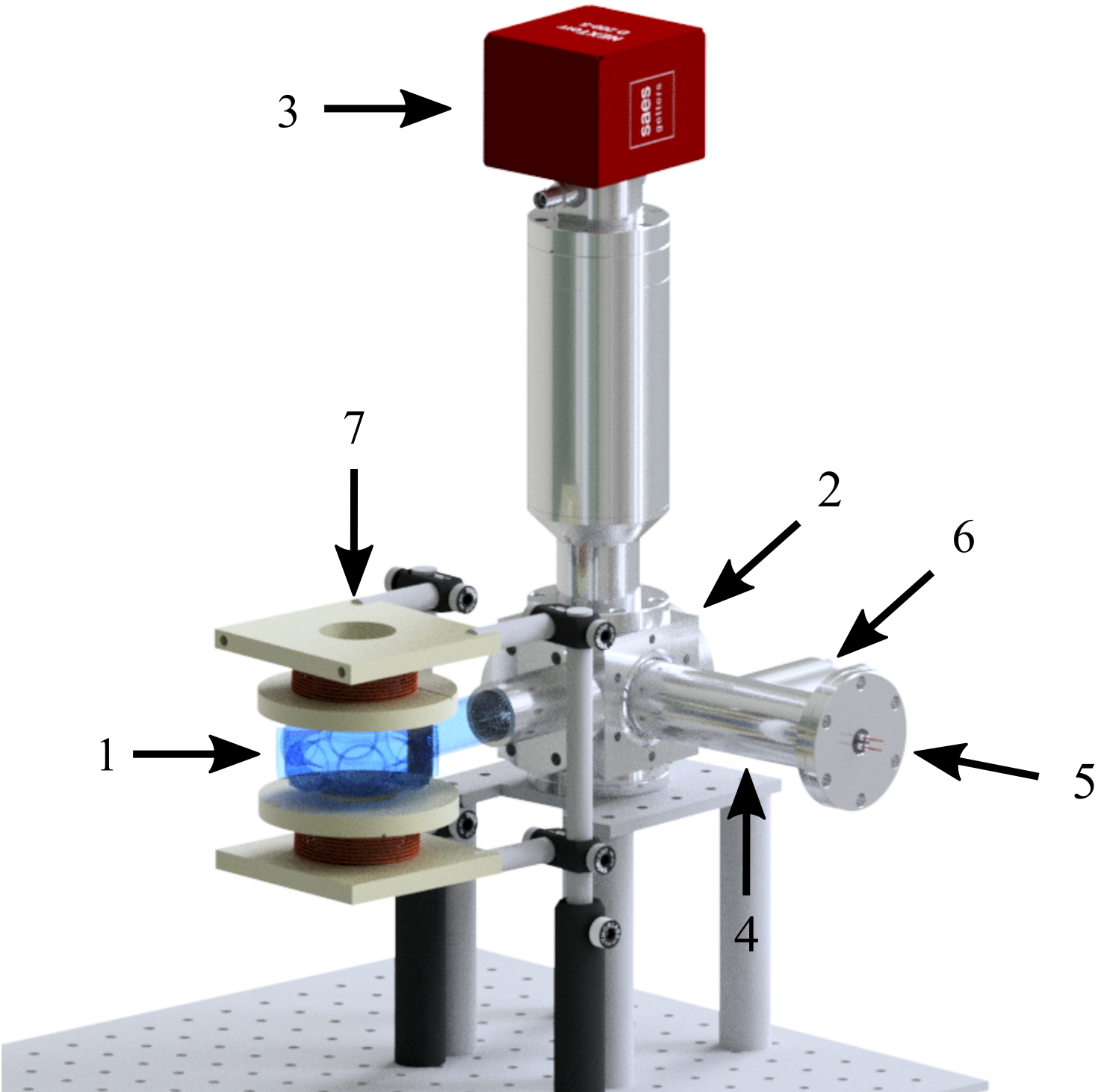}
    \captionof{figure}{Render of our vacuum system. The components are: (1) science cell, (2) CF30 cube, (3) vacuum pump (NEXTorr D 100-5), (4) tee, (5) feedthrough for Rubidium dispensers, (6) UHV valve and (7) MOT coils.}\label{fig:vacuum}
\endgroup

Figure \ref{fig:vacuum} also shows the MOT coils (7). They have 13 layers, with 9 windings each, of 13 AWG wire. The coil formers were 3D printed and designed to preserve full optical access to the science cell. Once arranged in an anti-Helmholtz configuration this coils generate a magnetic quadrupole field with axial gradients up to 30\,Gcm$^{-1}$, driving 8\,A, with a stable temperature under 60 ºC. We found easier to align the MOT and FWM beams for experiments without compensation coils. The three MOT beams (not shown) are delivered from their distribution board, depicted in Figure~\ref{fig:lasers}. They form a 3D retro-reflection laser cooling arrangement. Each beam is about 20 mm in diameter and carries  13\,mW of light that is - 24\,MHz detuned from the cooling transition. As explained below, less than 10 mW of evenly distributed re-pump light are enough to saturate the number of trapped atoms.

\subsubsection{Optical density}
\label{sub:Optical-density}

Collective quantum-optical phenomena depend upon the amount of emitting atoms. For the light generated via FWM this is determined by the number atoms simultaneously interacting with both pump beams. A direct way to find out this number is by measuring $OD$ in the atomic cloud. Together with its shape this parameter governs the influence that superradiance may have on the relaxation of atoms from their excited states and, consequently, on the bi-photons coherence properties \cite{jen2012spectral}.

\begingroup
    \centering
    \includegraphics[width=1\columnwidth]{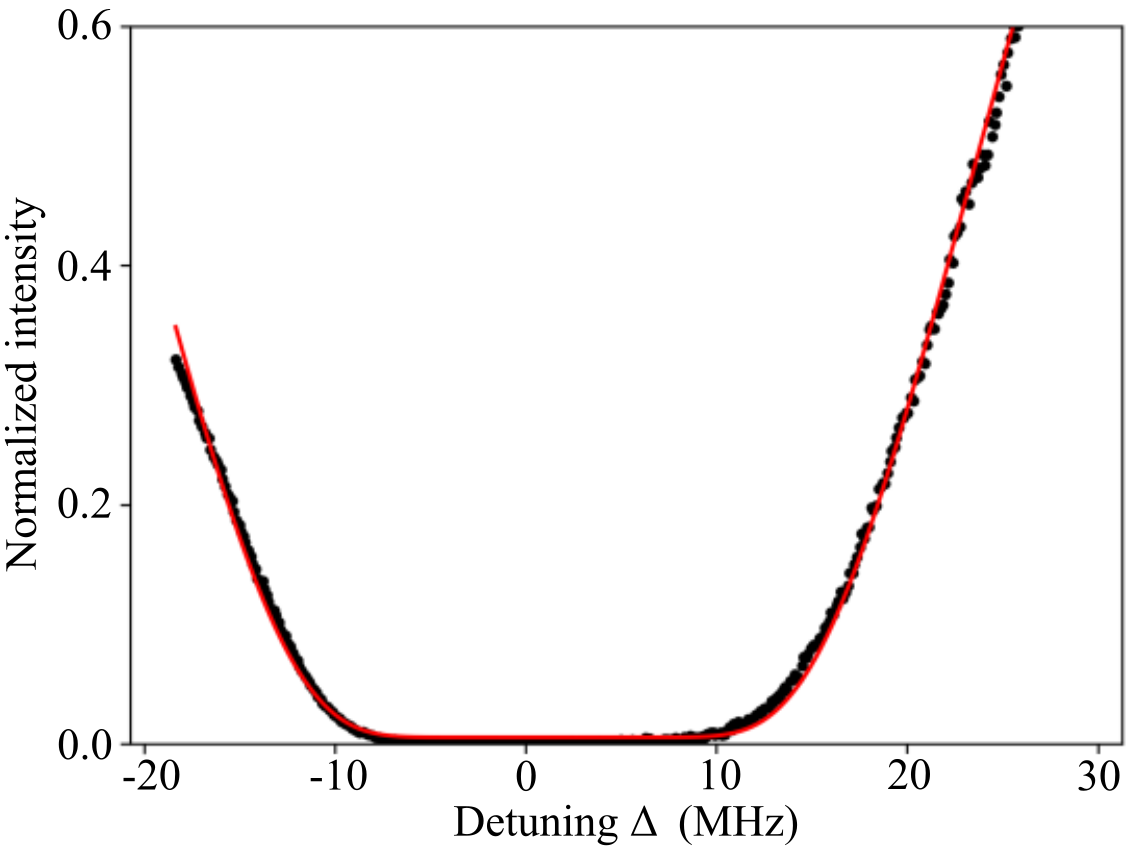}
    \captionof{figure}{Normalized intensity $I/I_0$ as a function of the probe beam detuning. In black is the experimental data and  the red curve is a fit using  Equation~\eqref{OD_Final}.}
    \label{fig:OD}
\endgroup

The absorption of a near-resonance laser beam traveling through an atomic sample is described by the Beer's Law,  
\begin{equation}\label{eq:dIntensity}
\dfrac{dI}{dz}=-\hbar \omega \gamma_{p} n,
\end{equation}
where $I$ is its intensity,  $z$ is the propagation distance through the atomic sample, $n$ is the density of atoms, $\omega$ is the angular frequency of the transition being probed and $\gamma_{p}$ is the total scattering rate  given by \cite{Metcalf2002cooling}
\begin{equation}
   \gamma_{p}=\dfrac{s_{0}\Gamma/2}{1+s_{0}+(2\Delta/\Gamma)^{2}}.
\label{Scatering_rate}
\end{equation}
In Equation~\eqref{Scatering_rate} $\Gamma$ is the excited state natural decay rate; $\Delta=\omega_{l}-\omega$ is the laser detuning from resonance; $s_{0}=I/I_{sat}$ and  $I_{sat}$ are the on-resonance saturation parameter and intensity, respectively. For a low saturation parameter ($I \ll I_{sat}$) Equation~ \eqref{Scatering_rate} simplifies to  
\begin{equation}
   \gamma_{p}=\dfrac{I\Gamma}{2I_{sat}}\dfrac{\Gamma^{2}}{\Gamma^{2}+4\Delta^{2}}
\label{Scatering_rate_simplified}.
\end{equation}
Substituting Equation~\eqref{Scatering_rate_simplified} into Equation~\eqref{eq:dIntensity} and integrating along $z$  gives rise to the ratio of intensity absorption
\begin{equation}
   \dfrac{I}{I_{0}}= \exp \left( -OD\dfrac{\Gamma^{2}}{\Gamma^{2}+4\Delta^{2}}\right),
\label{OD_Final}
\end{equation}
where  $I_0$ is the intensity of the probe beam before been scattered by atoms. The optical density $OD$ is defined in terms of the on-resonance cross section $\sigma_{0}$, the atom number $N$, and the light mode transverse area $A$ as $OD=N \sigma_{0} /A$.

To measure $OD$ we target the center of the atomic cloud with a low intensity probe beam, scan its frequency from - 18 MHz to + 25 MHz around the cooling transition and monitor its absorption with a photodiode. We acquire three series of data: (i) with the MOT and probe beam on, (ii) without the MOT and the probe on, and (iii) with the MOT on and probe switched off. Figure \ref{fig:OD} displays a typical absorption data set (black dots) and curve (red) fitted to it with Equation~(\ref{OD_Final}). There,  $\Gamma$ is set to 6.065 MHz \cite{Steck2003Rubidium} and $OD$ is left as the only free parameter. The experimental points displayed on Figure~\ref{fig:OD} were obtained with a $I_{0}=$ 0.5 $\textrm{mW} \textrm{cm}^{-2}$, horizontally polarized probe beam. The asymmetry around resonance is originated by the Zeeman sub-levels that are involved in the probed transition. From the fit we can estimate the atom number. For this experiments the probe beam is around 0.008 cm$^{2}$ and $\sigma_{0}=2.907\times10^{-9}$ cm$^{2}$ for vertically polarized light. Thus, in this case $OD = 20 \pm 0.2$, which is equivalent to $5\times10^{7}$ absorbing atoms.

 The optical density of a MOT mostly depends on: (i) the detuning $\varepsilon$ of its cooling light, (ii) the pressure inside the vacuum chamber, (iii) the power of the re-pump beam and, (iv) the gradient of magnetic field given by its coils. Figure \ref{fig:OD_parameters} (a) displays the $OD$ of our MOT as a function of the gradient (black dots). To get these data the other parameters were fixed to values established during a previous optimization: $\varepsilon= $ - 23 MHz,  a pressure of 10$^{-9}$ torr and (iii) 8 mW of re-pump power evenly distributed among the MOT beams. The blue dots in Figure~\ref{fig:OD_parameters} (a) were taken to characterize $OD$ as a function of the re-pump power. All parameters were kept at the mentioned values except the magnetic gradient which was left saturated at 22 Gauss cm$^{-1}$. In this case $OD$ increases approximately linearly up to about 8 mW, which was the power left on for the next data set. 
 
 The blue dots in Figure~\ref{fig:OD_parameters} (b) illustrate $OD$ as a function of $\varepsilon$. Note that, as well as with the re-pump power, this dependency is approximately linear dependency down to about - 23 MHz.  Finally, the black dots in Figure~\ref{fig:OD_parameters} (b) were taken with this detuning. They clearly show  non-linear behaviour of $OD$ as a function of pressure. 

From the data shown in Figure~\ref{fig:OD_parameters}  one can conclude that $OD$ can be varied from 9 to 22 in our setup given that the MOT is fully loaded. Most of this range is conveniently covered by the detuning $\varepsilon$ as the knob. The rest of it may be completed by using the dependency of $OD$ on the re-pump power.

\begingroup
    \centering
    \includegraphics[width=0.9\columnwidth]{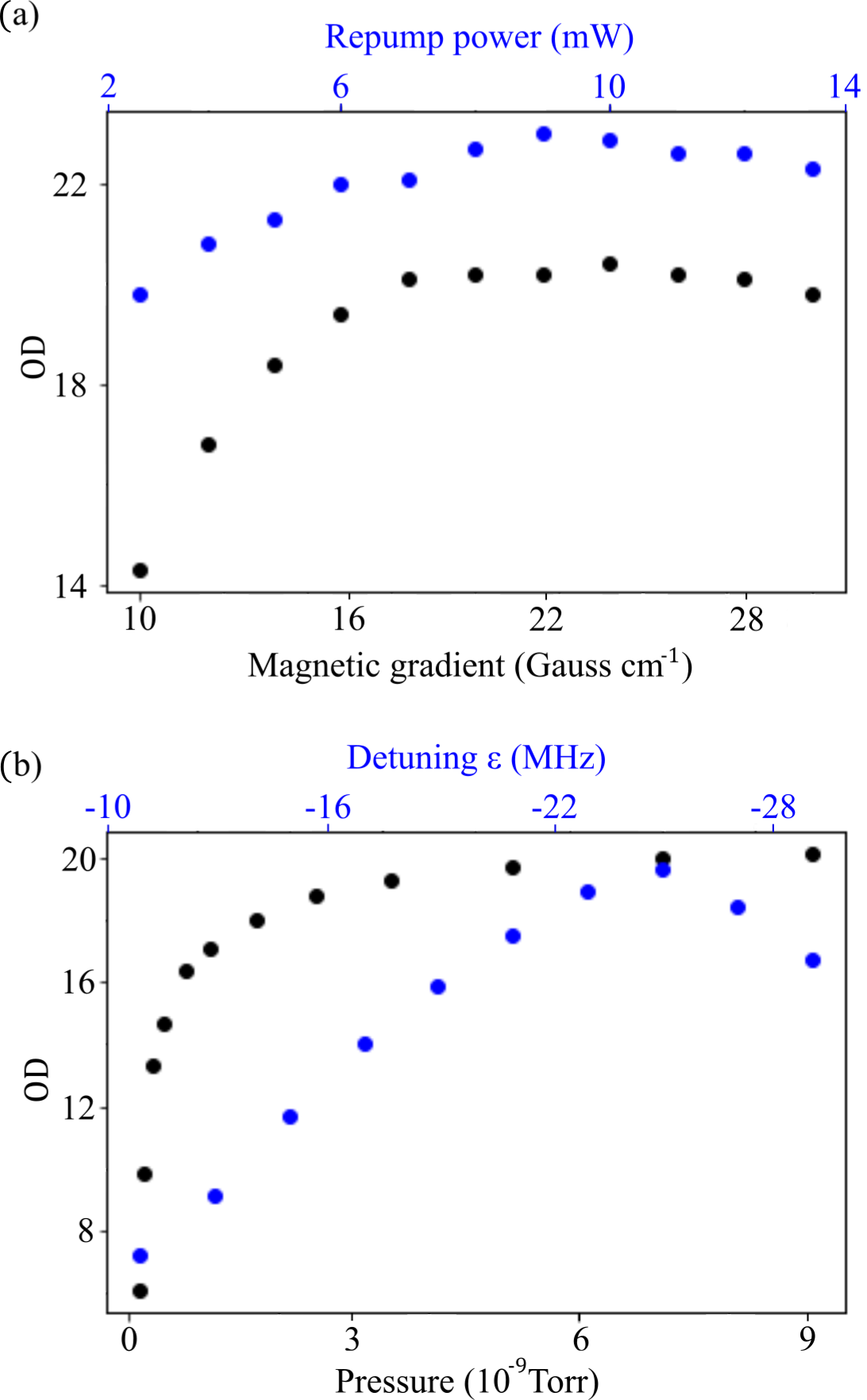}
    \captionof{figure}{Dependency of $OD$ on (a)  the power of the re-pump beam (blue) and on the field gradient produced by the coils (black) and; (b) on the detuning $\varepsilon$ (blue) and on the pressure inside the vacuum chamber (black).}\label{fig:OD_parameters}
\endgroup

\subsection{Control and DAQ}
\label{sub:control}

Many atomic physics laboratories around the world have developed their own control system according to their experimental needs. One can find online examples ranging from completely open source \cite{Laboratory_control} to semi-open source \cite{Board} and  closed systems \cite{Sinara}. They all have two common fundamental characteristics: parallelism in the input/output signals and the ability to transport high speed 50 $\Omega$ signals. The former is achieved with a FPGA or a DSP core (or DAQ card) and the latter with a high-speed buffer. 

In order to design our control system we thought of two generically different experimental sequences: (1) The most relevant parameter of the MOT to our current research interests is $OD$. We measure it as described in Section \ref{sub:MOT}. This method requires two analog input/output channels with no major speed demand. (2) To  generate photon pairs we need a duty cycle alternating MOT loading periods and FWM pulses, both in the order of hundreds of microseconds. As explained in Section \ref{sub:pumping} this can be achieved with 3 or 4 digital channels. 

To achieve parallelism the core must have the complete sequence information before outputting any signal. For this to be possible it requires to access a physical memory device (RAM or ROM), storing the control sequence previously sent by the control host, a PC in our case, see Figure~\ref{fig:control}. The time needed to deliver  information from the host to the core, usually referred to as communication time, must be considered if multiple sequences are to be exported. It is typically in the order of hundreds of milliseconds; normally the core takes a few nanoseconds to access the device memory. This time should also be taken into account for the duty cycle repetition accuracy. Based on the number of digital and analog channels, and speed demands to carry out experiments, we selected the LabJack T7 \cite{Labjack} as the core of our control system.

\subsubsection{Using a LabJack T7}

The LabJack T7 is an inexpensive DAQ card with 23 digital input/output channels, 2 analog outputs, and a 32 kB RAM memory card. It can be an good solution if its technical limits are well understood. Throughout developing this control system we tested its information management capacity and its time resolution.     

The digital channels are driven by 23 registers of one bit each\footnote{A register serves the sole purpose of storing the output value until the next one arrives. Therefore a time delay between the arrival of the information and its availability has to be taken into account.}; each analog output is driven by a different 32-bit register. It is worth mentioning that an analog output is created by using a digital to analog converter (DAC). Therefore, a binary number represents a voltage. The possible values of the analog outputs and their resolution depend on the length of the register assigned to it. The later is calculated by dividing the full span of the output voltage (3.3 V) by the complete number of combinations given the number of registers, 2$^{32}$ in this case.

The LabJack's RAM memory has 16-bit outputs (words) \cite{Labjack-stream}, therefore its core must create information packets of this size. In consequence all output registers (digital and analog) are connected to the RAM through a 16-bit buffer. Thus, all control sequences must be created using a 16-bit word convention. For example, if we had four 16-bit words in the RAM memory the first word could be used to drive the first 16 digital outputs (considering there are in total 23 digital outputs), the second word could drive the remaining 7 outputs (ignoring the remaining 9-bits), and the last two words would drive an analog output (given the fact that each analog output have a 32-bit resolution, it requires two 16 bit RAM words to be completely defined). Every access to a different RAM word has to be done in a different time slot, therefore only a 16-bit parallelism is achievable in the LabJack.

The time resolution of the Labjack's internal clock is 10\,$\mu$s. This is the interval between accessing a memory and setting its information into an output register. In \textit{stream mode} -- the fastest to transfer information -- the transmission frequency is 50 kHz~\cite{Labjack-stream}. Meaning that the complete sequence will be based on 20 $\mu$s time slots and its length will be limited by the RAM memory. Considering that 1 byte is equal to 8 bits and the 16-bit word convention, the maximum number of words is 16,384. The T7 allows either to work with 32-bits words in the case of analog outputs or to use the complete span over the digital outputs. In that case the maximum is reduced to 8,192 words. Thus, if we want to work with the minimum time scale and 16-bit RAM words, the LabJack would output all the data in 327.680 ms. This is more than enough for our current experimental needs. However, sequences requiring longer duty cycles -- e.g. absorption imaging of cold atoms -- may not be possible with the T7. If an analog output is used the resolution time is doubled because it needs 2 words per time slot to transmit the information; using the complete resources of the LabJack requires 4 words per time slot which is equal to 80\,$\mu$s.

\begingroup
    \centering
    \includegraphics[width=\columnwidth]{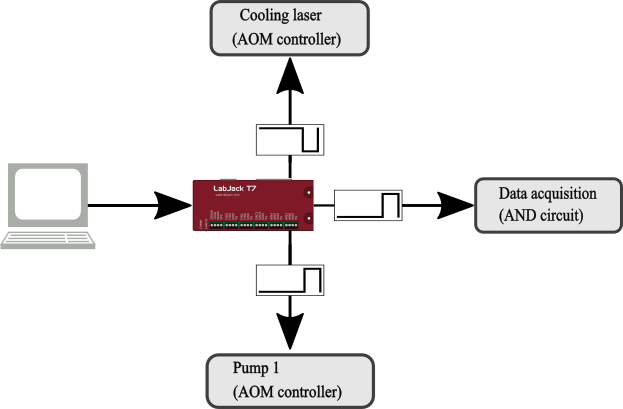}
    \captionof{figure}{Block diagram of our control system. The experimental sequences are programmed in a PC. This information is sent to the chosen core, a LabJack T7 card, which in turn sends the coordinating signals to the experimental devices and the data acquisition system, see Section \ref{sec:data_acq}.    }\label{fig:control}
\endgroup

The T7 has a D-sub output that we used for wiring to the experimental setup by building a BNC terminal board \cite{doi_rodrigo}. Its output currents are  low for driving 50 Ohm loads. Thus we built this board with a high speed buffer (Texas Instruments LMH6559) embedded on each channel. A two-rail power supply was used to bias the output buffers. The standby current consumption of all the buffered outputs is 500 mA. We wrote in Python a software to communicate a PC with the DAQ card. Its code and libraries are based on the vendors basic language, they are available in \cite{doi_rodrigo}.

\subsubsection{Data acquisition}
\label{sec:data_acq}

Data acquisition for this experiment is also driven by a combination of commercial and home made instruments and software. It uses up to four avalanche photodiodes (APD, ID120-500-800) gated by AND circuits to a time-to-digital converter (id800TDC). The generated information is handled by an application written in Python. 
All APDs were calibrated through their bias voltages and operating temperatures. The quantum efficiency was set equal for all of them by adjusting individually the bias voltage whilst detecting a specific light intensity and wavelength. The operating temperatures were adjusted to -40 $^{\circ}$C to reduce the dark count rate (DCR) due to thermically-generated carriers in the detectors below 200 Hz.
The data acquisition and control systems are synchronised through a home built AND circuit using a tri-state buffer (Texas Instruments SN74LVC126A) with a rising time of $<$10 ns. This circuit serves for gating the output from our APDs to the id800TDC; our specific APD model is free-running and has no internal trigger mechanism. The APDs and the id800TDC operate with LVTTL signals while the LabJack T7 uses TTL pulses: this requires the AND circuit to down convert everything to LVTTL, acting as an interface between the two systems. More details about this AND circuit are found in Appendix \ref{app:a}.
One of the main constraints to design our data acquisition system was the software proprietary set by the manufacturer of the time tagger. This software would only run on 32-bit Windows, which was inefficient with our setup\footnote{According to the vendor the newer model provides native support for Python but we were unable to find information on the compatibility with other operating systems.}.  An in-house solution was developed in Python as a library for interfacing with the id800TDC. This library presents many advantages over the original software: flexibility of use, possibility of automation, and the modularity required for its integration with the rest of our software ecosystem. A graphical interface was also developed for calibration and real-time visualization purposes. Our software also allows processing of data outputted from the id800TDC to the computer in either batching or streaming modes depending on the memory load of the application and the extension of the experiment. This is possible because the id800TDC has data transfer rates up to 2.5 million events per second. For experiments in which a large number of photon events might cause memory issues the streaming mode allows the information to be directly written to a file. All the necessary code to setup a similar system is available in \cite{Yves:2020}.

\
\end{multicols}
\begingroup
    \centering
    \includegraphics[width=0.85\textwidth]{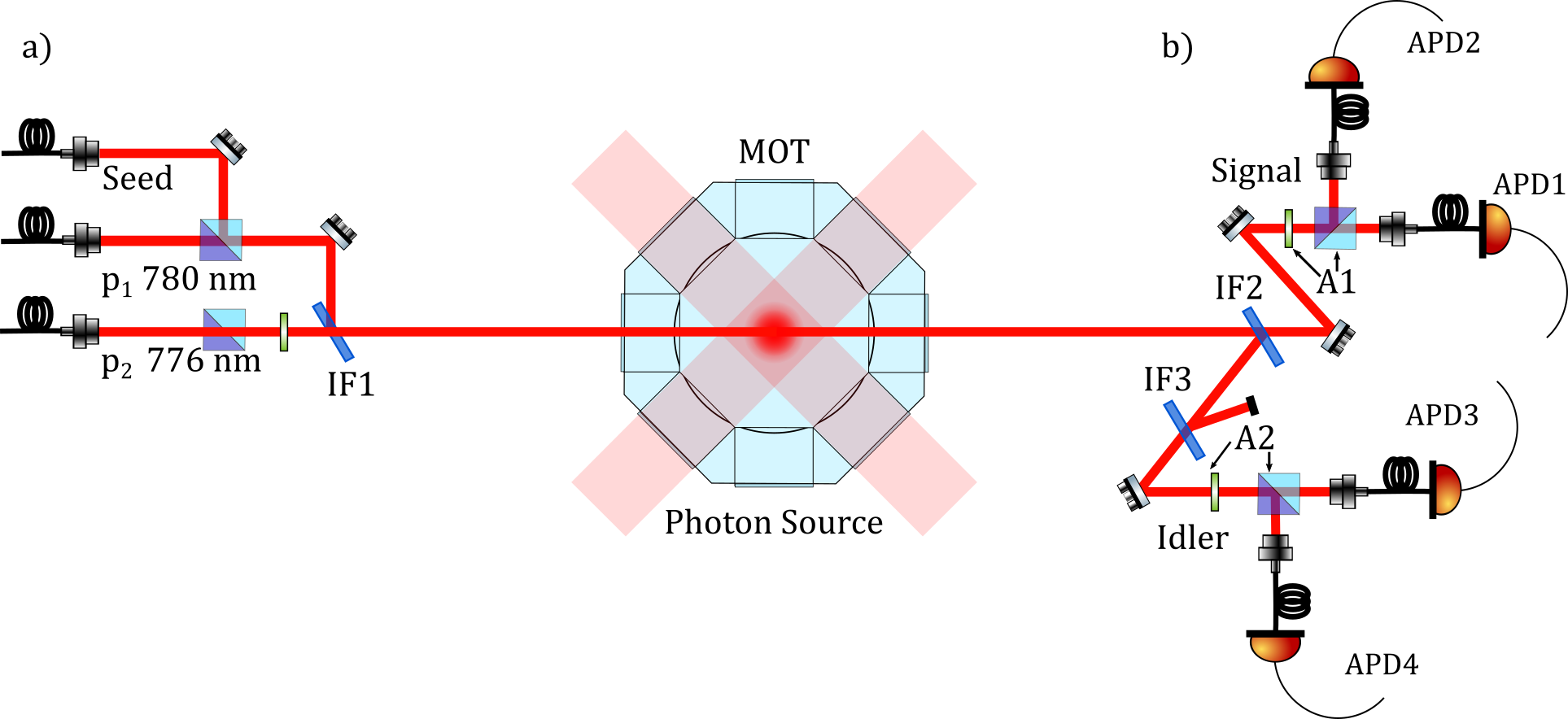}
    \captionof{figure}{ Schematics of the experimental setup to measure the time statistics of the photon pairs and their correlations in polarization: (a) is the optics arrangement to collimate and combine the pump beams toward the center of the MOT and; (b) is the optics for photon collection. The interference filter IF2 just transmits light with the signal frequency whereas IF3 transmits just idler photons and reflects all the pump beams remainders. Both output ports A1 and A2 couple each polarization component of the generated photons to an APD.     }
    \label{fig:setupExp}
\endgroup
\begin{multicols}{2}
\

\subsection{Pumping and detection}
\label{sub:pumping}

Four-wave mixing happens inside the science chamber that {\color{red} is} located at the center of the schematics on  Figure~\ref{fig:setupExp}, where the arrangements for the final preparation of the pump light and for the collection of the photon pairs are also depicted.

The three FWM beams, $p1$, $p2$ and  the seed, are delivered to their distribution board, Figure~\ref{fig:setupExp} (a). There, the three of them are collimated to a diameter of 1.1 mm with aspheric lenses (C230TMD-B) and are  combined on the narrow-band interference filter IF1 (LL01-780-25)  that transmits the 776 nm beam whilst reflecting light at the other two wavelengths. For the experiments reported here all polarizations were set linear. The three beams are overlapped in an optical path targeting the center of the atomic cloud. With this propagation geometry the mode-matching condition --  Equation~(\ref{eq:k_conservation}) -- is satisfied if the signal and the idler photons are generated in the same direction.

Pumping FWM in a co-linear configuration facilitates collecting the photon pairs given that one is able to separate the different frequencies from the four-coloured beam. In our setup this is done with two additional interference filters as shown in Figure~\ref{fig:setupExp} (b): IF2 (LL01-780-25) allows just photons with the signal wavelength to pass through; IF3 (LL01-808-25) transmits the idler photon wavelength only. This array also filters  the remainders of pump light in the photon flux that would appear as noise in our data. As depicted by Figure~\ref{fig:setupExp} (b) both the signal and idler channels (A1 and A2) consist on a $\lambda/2$ and PBS arrangement  enabling controllably spliting each photon flux in two inputs of fiber optics leading to the corresponding pair of avalanche detectors. 

Light sources at 762 nm and 795 nm are required to align the signal and idler photons because both channels are filtered in frequency. The two outputs of the signal channel A1 are aligned by enhancing the FWM process with the seed beam \cite{willis2009four}. This enables detecting 762 nm photons with a simple CMOS camera, which serves as the reference to couple these photons to the fibers of APD1 and APD2. Fibers leading the idler photons to APD3 and APD4 are aligned with the seed laser beam itself as the reference. Thus we typically reach a 70$\%$ fiber coupling efficiency using aspherical lenses  (C230TMD-B). Our APDs have a nominal quantum efficiency close to 80 \% around 780 nm. 

The optical setup illustrated in Figure~\ref{fig:setupExp} (b) has been designed for measuring the signal and iddler crossed correlations and the auto-correlations of each channel. Bi-photon correlations are measured by adjusting the $\lambda/2$ wave-plate for delivering all the light of each channel to a single APD. For auto-correlation experiments one can set each channel to split evenly its photon flux among the two detectors. Measurements of polarization correlations can be carried out as well since each APD is setup to measure just one component the generated photons.

A number of figures of merit have been designed to evaluate the generation of photon pairs. For this work we chose the coincidence rate since it is important for increasing the correlation histogram and the spectral brightness of the photon source (see Section \ref{sec:Statistical_measurements}).  This parameter is optimized by a detuning $\Delta=- 70$ MHz and a power of 500 $\mu$W  for $p1$ and,  a two-photon detuning $\delta =-3$ MHz and a 7 mW power for $p2$ in our setup; the FWM process is enhanced by setting the polarization of both pump beams linear and orthogonal to each other \cite{cere2018characterization}. Figure \ref{fig:timing} depicts the experimental duty cycle maximising the bi-photon coincidence rate. It consists on a loading time of 500 $\mu$s followed by a 200 $\mu$s pulse of FWM, when $p1$ and the AND circuits are switched on. Data acquisition is carried out during this period, without cooling light. The re-pump laser is left on at all times to maximize the number of atoms taking part in the double excitation scheme. The $p2$ beam is also switched on throughout the full duty cycle. This might be  limiting our photon generation by expelling a few atoms from the MOT.  We also leave the MOT coils switched on during the integration time. They generate a field around one Gauss at the atom-light interaction volume, close to the center of the MOT. This means that atoms are subject to a Zeeman shift in the order of 1.4 MHz during experiments. Our pump lasers have a bandwidth smaller than 250 kHz. We did not notice any effect of this magnetic field over the coherence results presented in Section \ref{sec:Statistical_measurements}. However, we might observe some of its influence over quantum polarization correlations during studies in the near future.

\
\begingroup
    \centering
    \includegraphics[width=0.8\columnwidth]{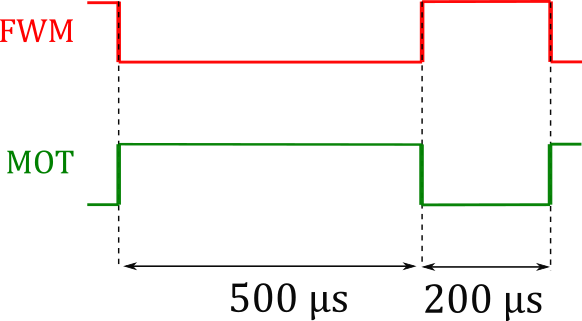}
    \captionof{figure}{Experimental duty cycle optimizing the coincidence rate of photon pairs in our setup. It consists on a 500 $\mu$s MOT loading time (green) followed by a 200 $\mu$s FWM pulse (red). During the FWM stage the cooling light is switched off at the same time of switching on the pump beams and the AND gate. The re-pump and $p2$ beams are always switched on as well as the MOT coils. }
    \label{fig:timing}
\endgroup
\

\section{Measured coherence properties }
\label{sec:Statistical_measurements}

With the setup described above we recorded the detection statistics of the photon pairs. Its behaviour is predicted by their second-order crossed-correlation function (which was derived in Section~\ref{sec:theory_coherence}), and the auto-correlation function of each channel. With the former we were able to estimate the coherence time of the heralded idler photons and thus, the spectral brightness of our source.  A brief theoretical discussion of the expected behavior of the auto-correlation functions $G^{(2)}_{ii}$ and $G^{(2)}_{ss}$ is given below. By measuring them it is confirmed that each decay channel corresponds to a chaotic source of photons, whose coherence time as such can be experimentally determined. Both results yield the required information to evaluate the classicality of the bi-photon temporal envelope through the Cauchy-Schwartz inequality.

\subsection{ Crossed- correlation function: coherence of the heralded photons}

From Equation (\ref{eq:G_2_double_decay}) is reasonable to assume that the measured crossed-correlation function should fit  
\begin{equation}
G^{(2)}_{si}(\Delta t)=G_{acc}+G_{0}\exp(-\Delta t/\tau_{c})\Theta(\Delta t),
\label{eq:G_cross}
\end{equation}
where $G_{0}$ is the maximum coincidence rate and 
\begin{equation}
    \label{eq:Gacc}
    G_{acc}=R_{1}R_{2}\Delta t_{\textrm{bin}}T
\end{equation}
is the number of accidental coincidences. It is given by the coincidence rate expected from two uncorrelated random sources; $R_{1}$ and $R_{2}$ are the individual count rates measured at each APD; $\Delta t_{\textrm{bin}}$ is the temporal bin width of the cross-correlation histogram and $T$ the total integration time.

We fit the experimental data to the degree of second order coherence, which is obtained by normalizing Equation~(\ref{eq:G_cross}) 
\begin{equation}\label{eq:g_cross}
g^{(2)}_{si}(\Delta t) = \frac{G^{(2)}_{si}(\Delta t)}{G_{acc}}.    
\end{equation}
To make a more realistic model, one needs to introduce the noise probability distribution $f(\Delta t)$ of each APD. Assuming that Equation~(\ref{eq:g_cross}) and $f(\Delta t)$ are two independent stochastic distributions, this measurement is expected to be a convolution of both of them. It is in general non-trivial to determine the function $f(\Delta t)$ of a detector. However, we found reasonable results by assuming that it is a Gaussian, 
 \begin{equation}\label{eq:Gaussian_noise}
     f(\Delta t) \propto \frac{1}{\tau_D \sqrt{2\pi}}\exp{\left(-\frac{\Delta t^2}{2\tau_D^2}\right)},
 \end{equation}
 with a width $\tau_D$ that is related to the response time of the detectors. Carrying out the convolution results in the fitting function
\begin{align}
 g^{(2)}_{si(fit)}(\Delta t)&=\frac{G_{0}}{2G_{acc}} \exp{\left(\frac{\tau_D^2-2\Delta t\tau_{c}}{2\tau_{c}^2}\right)}\\
&\times \left[ \erf \left(\frac{\Delta t\tau_{c}/\tau_D-\tau_D}{\sqrt{2}\tau_{c}}\right)+1\right].
\label{eq:fit_conv_crossed}
\end{align}
 
Coincidence histograms are obtained as explained in Section \ref{sec:setup}. Since FWM is pumped with two beams with orthogonal, linear polarization the signal beam is measured on transmission and the idler beam is measured on reflection from their corresponding PBS (Figure~\ref{fig:setupExp}). In this way the accidental coincidences are attenuated by approximately one-third. For the experimental results here presented the parameters of the pump beams are those given in Section \ref{sub:pumping} except that, in this case, the two-photon detuning $\delta$ = + 6 MHz. The optical density is set to approximately 20, T = 17 s and $\Delta t_{\textrm{bins}}=1.4$ ns.

Figure \ref{fig:g2_cross} shows the experimental data (black dots) for typical cross-correlation experiments in our setup. The dotted red curve is a fitting of Equation~(\ref{eq:fit_conv_crossed}) to these results leaving $G_0$, $\tau_D$ and $\tau_c$ as free parameters. One can immediately observe that the rise to its maximum value is smoother than the Heaviside function expected for ideal detection. The time dependency of these measurements is attenuated by the Gaussian noise distribution with bandwidth $\tau_D$. The fitted curve is optimum for $\tau_D \sim 0.61$ ns, which is in the order of our detectors nominal response time (400 ps). From the exponential decay of the fitted curve we measured a heralded coherence time of $\tau_{c}=4.41\pm 0.09$ ns for the idler photons. In a purely atomic cascade decay this would be expected to be equal to 27 ns, the lifetime of the intermediate state 5P$_{1/2}$. The observed reduction is consistent with collective effects expected in atomic ensembles \cite{nicholas1971superradiance}.

\
\begingroup
    \centering
    \includegraphics[width=\columnwidth]{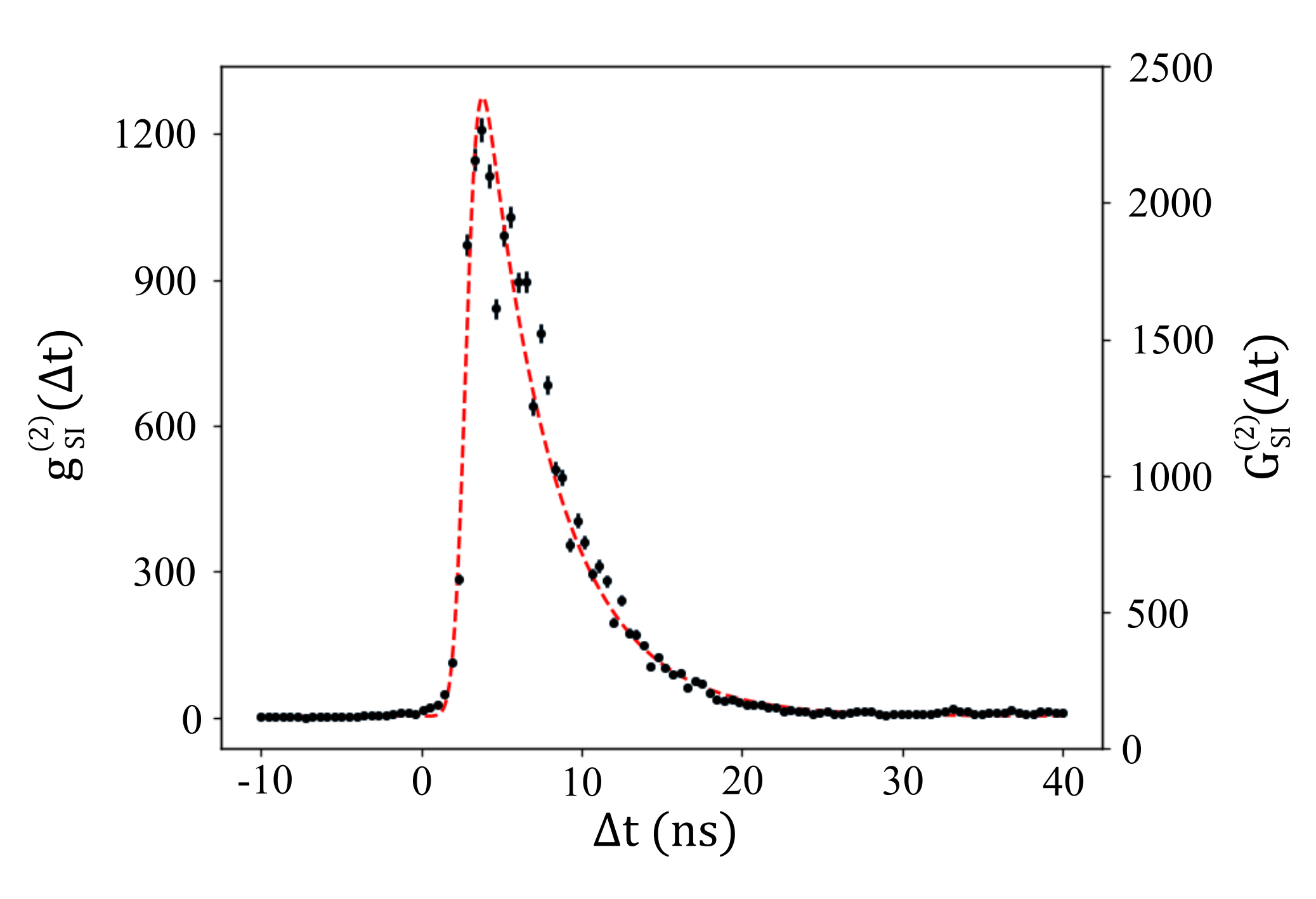}
    \captionof{figure}{ Cross-correlation histogram (black dots) as a function of delay the time between the signal and the idler photons. 
    The right axis scales for the raw data whereas the left displays values for the normalized histogram. The red curve is a fit with Equation~(\ref{eq:fit_conv_crossed}). The best fitting parameters are $G_{0}= 1654\pm 48$, $\tau_D=0.61\pm 0.04$ ns and $\tau_{c}=4.4\pm 0.1$ ns.}
    \label{fig:g2_cross}
\endgroup
\
We use two separate estimates of the accidental coincidences $G_{acc}$: either by their direct calculation, with Equation~(\ref{eq:Gacc}), or by fitting the experimental data with Equation~ (\ref{eq:G_cross}).
Calculation from the measured rates, $R_{1}=16,295$ $\textrm{s}^{-1}$ and $R_{2}=15,860$ $\textrm{s}^{-1}$, and the time settings for these experiments yields $G_{acc} \sim 5.6$ $\textrm{s}^{-1}$.
In order to obtain a reliable estimate from Equation~(\ref{eq:Gacc}), we consider the longer coincidence time interval of  $\Delta t= 350$ ns, resulting in $G_{acc}=5.8 \pm 1.1$, compatible with the statistically inferred value.


The measured coherence time of the heralded idler photons corresponds to a bandwidth of $36.2 \pm 0.8$ MHz, which is a factor of ten smaller than the value obtained in experiments without cooling the atoms \cite{willis2009four}. It in principle can be reduced to 20 MHz by optimizing the pumping parameters  \cite{srivathsan2013narrow}, and below 8 MHz by removing the time-dependent phases with a Fabry-Perot cavity and a electro-optic modulator \cite{Seidler:2020jm}. The 5S$_{1/2}\rightarrow$ S$_{1/2}$ transition of Rb$^{87}$ has a natural line with of 5.75 MHz~\cite{Steck2003Rubidium}. Thus heralded photons sourced in this way are suitable for interacting with atoms.

The spectral brightness of a quantum light source is likely to become a determining parameter to achieve control over the interaction of its photons with atoms. It is given by $B=2\pi \tau_{c}r_c$, where $2\pi \tau_{c}$ is the inverse of its single-photon bandwidth, 2$\pi$ (4.4 $\pm$ 0.1 ns) in our case. The coincidence rate $r_c$ is obtained from the un-normalized histogram $G^{2}_{SI}$ plotted in Figure~\ref{fig:g2_cross}. It is given by the number of coincidences detected during an integration window that we chose to be 40 ns. These yield a coincidence rate up to  $10^{4}$ s$^{-1}$ under experimental conditions similar to the reported above but with $\delta$ = 0.5 MHz. Thus we report here a spectral brightness around  280 coincidences s$^{-1}$ MHz. This value is consistent with observations in similar sources \cite{cere2018characterization} and two orders of magnitude larger than the value reported for hot atomic ensembles \cite{willis2009four}.

\subsection{Auto-correlation functions: coherence properties of each channel. }

Since the FWM process is induced on many atoms, each decay  channel is a source of photons with chaotic origin. One can find their coherence properties through the treatment shown in Section \ref{sub:heralding_effect}, or by recalling that the first order correlation function of each atomic relaxation is given by the Fourier transform of its Lorentzian spectrum  $S_{i}(\omega)=\Gamma_i/2[(\omega-\omega_i)^2-(\Gamma_i/2)^2]$~\cite{Carmichael:2013ut}; 
\begin{equation}
    G_{ii}^{(1)}(\Delta t)=\frac{1}{2\pi}\int_{-\infty}^{\infty}d\omega e^{i\omega\Delta t }S(\omega)=G_0e^{-i\omega_0\Delta t}e^{ - \Delta t/\tau_c},
\end{equation}
where $G_0 \simeq \sqrt{G_{acc}}$  and $\tau_c$ is the coherence time of the chaotic photons, signal or idler. Due to their nature, in both cases, the second order of coherence is related to normalized first order correlation
\begin{equation}
    g_{ii}^{(1)}(\Delta t)= \frac{G_{ii}^{(1)}(\Delta t)}{G_0}
\end{equation}
by 
\begin{equation}\label{eq:g2_auto}
    g_{ii}^{(2)}(\Delta t)=1+|g_{ii}^{(1)}(\Delta t)|^2 = 1+e^{-2|\Delta t|/\tau_c}, 
\end{equation}
indicating an expected sharp time symmetry due to the absolute value in the exponent with an exponential decay with rate $\Gamma = 1/(2\pi\tau_c)$ towards negative and positive times.

To fit this data, the detectors noise was also assumed Gaussian.  Its convolution with Equation~ (\ref{eq:g2_auto}) is 
\begin{align}
    g^{(2)}_{ii(fit)}(\tau) & =\frac{1}{2}g_{0} e^{\left(\frac{\tau_D^2-2\Delta t\tau_{c}}{2\tau_{c}^2}\right)} \times \nonumber \\
    & \left[ \erfc \left(\frac{\tau_D^2-\Delta t\tau_{c}}{\sqrt{2}\tau_D\tau_{c}}\right)+ e^{\frac{2\Delta t}{\tau_{c}}} \erfc \left(\frac{\tau_D^2+\Delta t\tau_{c}}{\sqrt{2}\tau_D\tau_{c}}\right) \right].
    \label{eq:fit_conv_auto}
\end{align}
 
 \
\begingroup
    \centering
    \includegraphics[width=0.95\columnwidth]{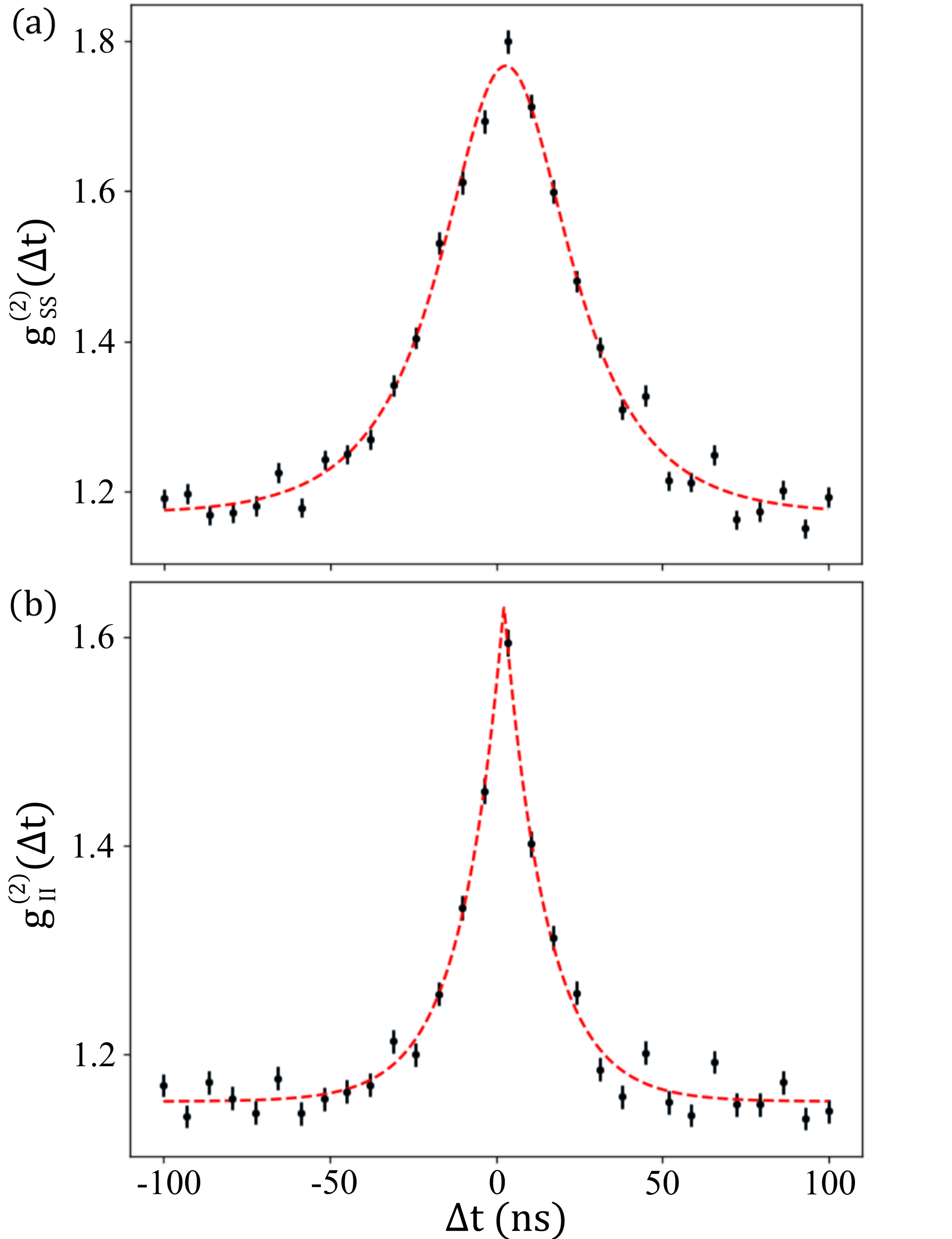}
    \captionof{figure}{ The black dots are the normalized autocorrelation histograms for (a) signal  and (b) the idler photons. Both red curves represent the corresponding fits according to Equation~ (\ref{eq:fit_conv_auto}). In (a) the best fitting parameters are $g_{0}=0.22 \pm 0.03$, $\tau_D=9.77\pm 2.12$ ns and $\tau_{c}=18.92\pm2.65$ ns; in (b) $g_{0}=0.12\pm 0.03$, $\tau_D=0.17\pm 0.04$ ns and $\tau_{c}=12.80\pm 0.90$ ns.} 
    \label{fig:auto}
\endgroup
\

These experiments were performed by implementing a Hanbury Brown–Twiss interferometer at each output channel, A1 and A2, depicted on Figure~\ref{fig:setupExp} (b). The photon flux of each one was evenly split at the PBS. Both reflecting and transmitting outputs were coupled into single-mode fibers and sent to their own APD.

The black dots in Figure~\ref{fig:auto} display typical coincidence histograms for the signal (a) and idler (b) photons; the dashed, red curves are fitted to these data with Equation~(\ref{eq:fit_conv_auto}).  All the experimental parameters were set the same as for the cross-correlation plots except that the two photon detunings were $\delta=$ 4 MHz and $\delta=$ 7 MHz for (a) and (b) respectively and, the integration time was 20 minutes in both cases. 

It is illustrative to point out the differences between these observations and the behaviour predicted for ideal detection. According to Equation~(\ref{eq:g2_auto}) the auto-correlation should present a maximum equal to 2 at the time zero. Neither the signal nor the idler maximum reach this value because the APDs quantum efficiencies are lower than 100\%. Furthermore, it was found that $g^{(2)}_{ss}(0) > g^{(2)}_{ii}(0)$, which is consistent with the greater quantum efficiency of our detectors at 762 nm than at 795 nm. Note that the experimental maxima have been smoothed by the noise bandwidth $\tau_D$, which was left as a free parameter during the fittings. Their values, displayed on Table \ref{tab:Table1}, are different because the noise of the detectors is a function of the optical wavelength. In this case $\tau_{D\mathrm{762}}>\tau_{D\mathrm{795}}$ translates to a softer maximum for the signal auto-correlation. 

The coherence time $\tau_c$ was left as a free parameter for these fits. Its values for the signal and idler photons are also displayed in Table \ref{tab:Table1}. They are consistently longer than the coherence time of the heralded idler photons, as observed in similar sources of photon pairs \cite{srivathsan2013narrow}.

\begin {table}[H]
 \begin{center}
 \scalebox{1.0}{
  \begin{tabular}{ | >{\centering\arraybackslash}m{0.60in}  | >{\centering\arraybackslash}m{.60in} | >{\centering\arraybackslash}m{.60in} | >{\centering\arraybackslash}m{.60in} |}
  \hline
  & {\bf $g^{(2)}_{SI}$} & {\bf $g^{(2)}_{SS}$ } & {\bf $g^{(2)}_{II}$} \\ \hline
   $\tau_D$ (ns)      & $0.61\pm0.04$     &  $9.77\pm 1.12$      &  $0.17\pm0.04$    \\ \hline
    $\tau_{c}$  (ns)      &   $4.4\pm0.1$    & $18.9\pm2.7$       &  $12.8\pm0.9$     \\ \hline
  \end {tabular}}
 \end{center}
 \caption{Fit parameters $\tau_D$ and $\tau_c$ to the experimental data obtained for the cross-correlation, and the auto-correlation of the signal  and idler photons; $g^{(2)}_{si}$, $g^{(2)}_{ss}$ and $g^{(2)}_{ii}$ respectively.} \label{tab:Table1}  
\end {table}

\subsection{Non-classicality }

A vector analysis of the classical electromagnetic field, based on the Cauchy-Schwartz inequality, predicts that  \cite{reid1986violations}
\begin{equation}\label{eq:cauchy}
    R= \frac{[g^{(2)}_{si}(\Delta t_{\textrm{max}})]^{2}}{g^{(2)}_{ss}(0)g^{(2)}_{ii}(0)} \leq 1,
\end{equation}
where $g^{(2)}_{si}(\Delta t_{\textrm{max}})^{2}$, $g^{(2)}_{ss}(0)$ and $g^{(2)}_{ii}(0)$ are the maximum values of the second order of coherence measured from cross-correlations and autocorrelations. Their values, extracted from curves fitted to the experimental data shown in Figs.~\ref{fig:g2_cross} and \ref{fig:auto}, are: $g^{(2)}_{si}(\Delta t_{\textrm{max}})=1270\pm 48$, $g^{(2)}_{ss}(\Delta t_{\textrm{max}})=1.77\pm 1.5$ and $g^{(2)}_{ii}(\Delta t_{\textrm{max}})=1.63 \pm 1.4$. By plugging them into Equation~(\ref{eq:cauchy}) one finds that $R=5.62\times10^{5}$, which is evidence of a strong  non-classical behavior in the time statistics of the photon pairs generated by our source.

\section{Conclusion and outlook}
\label{sec:conclusion_outlook}

This article presented our source of photon pairs generated from cold atomic Rubidium samples. The employed experimental methods and techniques were described in detail. The idler photons have a heralded coherence time of 4.4 $\pm$ 0.1 ns. Together with the observed pair generation rates, which are in the order of $10^4$ s$^{-1}$, this coherence time yields a spectral brightness three orders of magnitude larger than typical values achieved by sources based on hot atomic gases \cite{willis2009four}. 

The reported results are a basic coherence characterisation of the generated photons pairs. Further insight is possible by replacing the presented model, which is purely atomic and is based on the  Schrödinger equation, by a mathematical treatment of the Lindblad type. With the later is possible to systematically calculate  atomic collective effects on the generated light, and to describe the bi-photon states in terms of their variables, as polarization and angular momentum. This would give a suitable guide for further experimental research seeking to achieve control over the generated quantum states with the valuable help of atomic spectroscopy. 

Our source is bright and the generated light has a bandwidth suitable for interacting with atoms. Tailoring the bi-photon states would allow us to modulate this interaction with precision. The achievement of these goals would broaden the possibilities for building complete quantum systems of time-correlated photon pairs with atoms. One of several potential possibilities in such kind of system is to imprint and read memories using the bi-photons as the flying messengers of quantum information. This is the scientific pathway that we have chosen for building an experimental scenario where exciting research on quantum information and its applications, like telecommunications and cryptography, can take place.

\section{Acknowledgments}
This is the first experimental setup of its kind in Mexico and the second in Latinamerica \cite{Felinto:2015cl}. We wish that our work serves to encourage the activity on related research topics in the region. During the construction of our laboratory we have been strongly supported by many colleagues and several institutions. We  wish to express our gratitude to our colleagues from the Institute of Physics, UNAM, Manuel Torres Labansat, Roberto José Raúl Gleason Villagrán, Jorge Amin Seman Harutinian, Freddy Jackson Poveda Cuevas, Asaf Paris Mandoki and Carlos Villareal Luján.  We thank our collaborators José I. Jiménez Mier y Terán and Fernando Ramírez Martínez from the Institute of Nuclear Sciences, UNAM, for numerous technical and academic discussions. We thank Jorge. G. Acosta Montes, Diego Sierra Costa and Carlos Luis Hernández Cedillo,  former members of our Labortatory, who helped in its setup. Finally, a special thanks to Giorgio Colangelo for his enthusiastic input to our group on the quest for achieving the first MOT. 

Secondly, we want to deeply express our gratitude to  the institutions that have financed the lab construction and its maintenance. Firstly, we acknowledge CONACyT and CTIC-UNAM for providing the resources for refurbishing the laboratory space and acquiring its mayor equipment through the projects LN232652, LN260704, LN271322, LN280181, LN293471, LN299057 and 314860,  granted in the National Laboratories calls from 2014 until 2020. This support was complemented by other research grants providing an important part of the operating expenses: The Institute of Physics Internal Project PIIF-008 in its call from 20016 until 2019; the PAPIIT projects IA103216,  IN108018 and IN106821 from DGAPA-UNAM and the Basic-Science project no. 285387, SEP-CONACyT.

Y. M. T. acknowledges DGAPA-UNAM for its support through a postdoctoral grant.  N. A. T. acknowledges the London Mathematical Laboratory, SEP-CONACyT and for providing posdoctoral support. She also thanks  the IF-UNAM and the CTIC-UNAM, for extending this support during the current SARS-CoV-2 outbreak. I. F. A. A. and L. A. M. L. acknowledges CONACyT for their graduate scolarships.  

A. C. acknowledges the  several grants that made possible his frequent visits to our Institute. In 2015 his trip was financed by LANMAC-CONACyT and  IF-UNAM. His travel expenses in 2016 were supported by the grant PIIF-008. Finally, in 2018 he held the Angel Dacal visiting Chair awarded by the Institute of Physics at UNAM.

\end{multicols}

\medline

\begin{multicols}{2}

\bibliographystyle{apsrev4-1}

\begin{thebibliography}{}

\bibitem{burnham1970observation} 
D. C. Burnham and D. L. Weinberg, Physical Review Letters \textbf{25}, 84 (1970)

\bibitem{freedman1972experimental}
S. J. Freedman and J. F. Clauser, Physical Review Letters
\textbf{28}, 938 (1972). 

\bibitem{Dowling:2003eia}
J. P. Dowling and G. J. Milburn, Philosophical Transactions
of the Royal Society of London. Series A: Mathematical,
Physical and Engineering Sciences (2003).

\bibitem{Couteau:2018gv}
C. Couteau, Contemporary Physics \textbf{59}, 291 (2018).

\bibitem{aspect1982experimental}
A. Aspect, P. Grangier, and G. Roger, Physical Review
Letters \textbf{49}, 91 (1982).

\bibitem{Duan:2001dz}
L. M. Duan, M. Lukin, J. I. Cirac, and P. Zoller, Nature
\textbf{414}, 413 (2001).

\bibitem{cho2016highly}
Y.-W. Cho, G. Campbell, J. Everett, J. Bernu, D. Higginbottom,
M. Cao, J. Geng, N. Robins, P. Lam, and
B. Buchler, Optica \textbf{3}, 100 (2016).

\bibitem{guo2019high}
J. Guo, X. Feng, P. Yang, Z. Yu, L. Chen, C.-H. Yuan,
and W. Zhang, Nature Communications \textbf{10}, 1 (2019).

\bibitem{Wallquist:2009efa}
M. Wallquist, K. Hammerer, P. Rabl, M. Lukin, and
P. Zoller, Physica Scripta \textbf{2009}, 014001 (2009).

\bibitem{Kurizki:2015ew}
G. Kurizki, P. Bertet, Y. Kubo, K. Mølmer, D. Petrosyan,
P. Rabl, and J. Schmiedmayer, Proceedings of
the National Academy of Sciences of United States of
America \textbf{112}, 3866 (2015).

\bibitem{Kuzmich:2003fe}
A. Kuzmich, W. Bowen, A. Boozer, A. Boca, C. Chou,
L.-M. Duan, and H. Kimble, Nature \textbf{423}, 731 (2003).

\bibitem{Thompson:2006gf}
J. K. Thompson, J. Simon, H. Loh, and V. Vuleti´c, Science
\textbf{313}, 74 (2006).

\bibitem{Chaneliere:2006gx}
T. Chaneliere, D. Matsukevich, S. Jenkins, T. Kennedy,
M. Chapman, and A. Kuzmich, Physical Review Letters
\textbf{96}, 093604 (2006).

\bibitem{Du:2008do}
S. Du, P. Kolchin, C. Belthangady, G. Y. Yin, and S. E.
Harris, Physical Review Letters \textbf{100}, 183603 (2008).

\bibitem{srivathsan2013narrow}
B. Srivathsan, G. K. Gulati, B. Chng, G. Maslennikov,
D. Matsukevich, and C. Kurtsiefer, Physical Review
Letters \textbf{111}, 123602 (2013).

\bibitem{Seidler:2020jm}
M. A. Seidler, X. J. Yeo, A. Cerè, and C. Kurtsiefer,
Physical Review Letters \textbf{125}, 183603 (2020).

\bibitem{Willis:2011ta}
R. T. Willis, F. E. Becerra, L. A. Orozco, and S. L.
Rolston, Optics Express \textbf{19}, 14632 (2011).

\bibitem{Gulati:2015ee}
G. K. Gulati, B. Srivathsan, B. Chng, A. Cerè, and
C. Kurtsiefer, New Journal of Physics \textbf{17}, 093034
(2015).

\bibitem{Zhao:2019ix}
T.-M. Zhao, Y. S. Ihn, and Y.-H. Kim, Physical Review
Letters \textbf{122}, 123607 (2019).

\bibitem{wen2006transverse}
J.Wen and M. H. Rubin, Physical Review A \textbf{74}, 023808
(2006).

\bibitem{scully1999quantum}
M. O. Scully and M. S. Zubairy, “Quantum optics,”
(1999).

\bibitem{kocher1971time}
C. A. Kocher, Annals of Physics \textbf{65}, 1 (1971).

\bibitem{stroud1972superradiant}
C. Stroud Jr, J. Eberly, W. Lama, and L. Mandel, Physical
Review A \textbf{5}, 1094 (1972).

\bibitem{sahagun2013simple}
D. Sahagun, V. Bolpasi, and W. von Klitzing, Optics
Communications \textbf{290}, 110 (2013).

\bibitem{mandal2018blue}
P. K. Mandal, V. Naik, V. Dev, A. Chakrabarti, and
A. Ray, Applied Optics \textbf{57}, 3612 (2018).

\bibitem{grove1995two}
T. T. Grove, V. Sanchez-Villicana, B. Duncan,
S. Maleki, and P. Gould, Physica Scripta \textbf{52}, 271
(1995).

\bibitem{Harris:2006}
M. L. Harris, C. S. Adams, S. L. Cornish, I. C. McLeod,
E. Tarleton, and I. G. Hughes, Physical Review A \textbf{73},
062509 (2006).

\bibitem{thompson2012narrow}
D. J. Thompson and R. E. Scholten, Review of Scientific
Instruments \textbf{83}, 023107 (2012).

\bibitem{barger1969pressure}
R. Barger and J. Hall, Physical Review Letters \textbf{22}, 4
(1969).

\bibitem{jen2012spectral}
H. Jen, Journal of Physics B: Atomic, Molecular and
Optical Physics \textbf{45}, 165504 (2012).

\bibitem{Metcalf2002cooling}
H. J. Metcalf and P. V. der Straten,
Laser Cooling and Trapping (Springer, 2002).

\bibitem{Steck2003Rubidium}
D. A. Steck, \href{http://steck.us/alkalidata}{available online}
 (revision 2.2.2, 9 July 2021). 

\bibitem{Laboratory_control}
Atom Optics Laboratory, University of Texas at Austin,
\href{http://m-labs.hk/experiment-control/sinara-core/}{A Laboratory Control System for Cold Atom Experiments}
(2015, [accessed 2020 Aug 20]).

\bibitem{Board}
European Laboratory for Non-Linear Spectroscopy,
\href{https://ew.lens.unifi.it/.}{DIO128-Board Based Aquisition System}
 (2005, [accessed
2020 Aug 20]).

\bibitem{Sinara}
M. Senn, \href{http://m-labs.hk/experiment-control/sinara-core/}{Sinara hardware. 2020. M-Labs} ([accessed
2020 Aug 20]).

\bibitem{Labjack}
LabJack Corporation, \href{https://labjack.com/products/t7}{“T7,”} ([accessed 2021 March
24]).

\bibitem{Labjack-stream}
LabJack Corporation, \href{https://labjack.com/support/datasheets/t-series/communication/stream-mode/stream-out}{“Stream mode,”} ([accessed 2021
July 1]).

\bibitem{doi_rodrigo}
R. A. Gutierrez-Arenas, \href{https://doi.org/10.5281/zenodo.5062126}{“Labjack T7 BNC terminal board and code,”} (2021).

\bibitem{Yves:2020}
L. Y. Villegas-Aguilar, \href{https://doi.org/10.5281/zenodo.4299245}{“pyid800: A Python interface for the ID800 time-tagger,”} (2020).


\bibitem{willis2009four}
R. Willis, F. Becerra, L. Orozco, and S. Rolston, Physical
Review A \textbf{79}, 033814 (2009).

\bibitem{cere2018characterization}
A. Cerè, B. Srivathsan, G. K. Gulati, B. Chng, and
C. Kurtsiefer, Physical Review A \textbf{98}, 023835 (2018).

\bibitem{nicholas1971superradiance}
N. E. Rehler and J. H. Eberly, Physical Review A \textbf{3},
1735 (1971).

\bibitem{Carmichael:2013ut}
H. Carmichael, Statistical Methods in Quantum Optics
(Springer Science \& Business Media, 2013).

\bibitem{reid1986violations}
M. Reid and D. Walls, Physical Review A \textbf{34}, 1260
(1986).

\bibitem{Felinto:2015cl}
Felinto, D, Borba, G C, Tabosa, J W R, de Oliveira, R
A, Barreiro, S, and Martins, W S, Optics Letters \textbf{40},
4939 (2015).

\end{thebibliography}

\appendix

\section{AND circuit}
\label{app:a}
A schematics of the gating circuit used to interface the APDs and the id800TDC is shown in Figure~\ref{fig:gatting}. The SN74LVC126A is a high speed buffer gate. In order to achieve the required voltages at 50 $\Omega$ terminations it was connected in series with a LMH6559 high-speed closed-loop buffer amplifier. The SN74LVC126A interfaces directly with the control TTL signals from the BNC distribution board of the control system. A 3.3 V power supply with standard coupling filters was used for the entire circuit. Its gerber files are available in \cite{Yves:2020}.

One circuit is used for each APD. It has two entries, the signal and the control inputs. The signal input is connected to an APD and the control receives instructions from the Labjack T7. The control input gates the detectors signal for each measurement to be delivered to the id800TDC by the output of the circuit. 

\
\begingroup
    \centering
    \includegraphics[width=0.95\columnwidth]{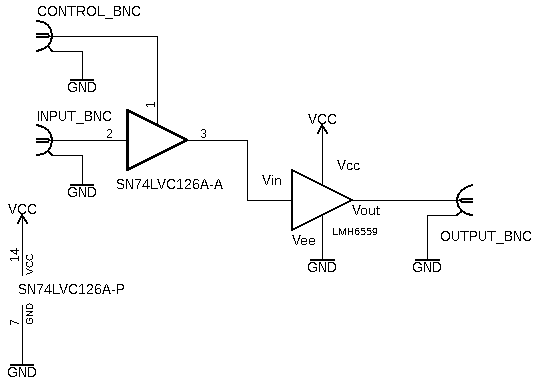}
    \captionof{figure}{Diagram of the AND circuit for gating the APD's. 
 The control inputs switches on and off the signal from the APD, which is delivered to the id800TDC for data acquisition.  } 
    \label{fig:gatting}
\endgroup
\
\end{multicols}
\end{document}